\documentclass{aa}  

\usepackage{graphicx}
\usepackage{txfonts}
\usepackage{color}
\newcommand{\NeII}{Ne\,{\sc ii}}
\newcommand{\OI}{O\,{\sc i}}

\begin{document}

   \title{High-speed molecular cloudlets around\\ the Galactic center’s supermassive black hole\thanks{This paper makes use of the following ALMA data: ADS/JAO.ALMA\#2013.1.01058.S.  ALMA is a partnership of ESO (representing its member states), NSF (USA), and NINS (Japan), together with NRC (Canada), and NSC and ASIAA (Taiwan), in cooperation with the Republic of Chile. The Joint ALMA Observatory is operated by ESO, AUI/NRAO, and NAOJ.}}

\titlerunning{Molecular cloudlets around Sgr~A$^*$} 
\authorrunning{J.R. Goicoechea et al.}

   \author{Javier R. Goicoechea\inst{1}
          \and
          Jerome Pety\inst{2,3}
          \and
          Edwige Chapillon\inst{2,4} 
          \and
          Jos\'e Cernicharo\inst{1}
          \and 
          Maryvonne Gerin\inst{3}
          \and\\
          Cinthya Herrera\inst{2} 
          \and
          Miguel A. Requena-Torres\inst{5}
          \and
          Miriam G. Santa-Maria\inst{1}
          }

   \institute{Instituto de F\'{\i}sica Fundamental (CSIC). Calle Serrano 121, 28006, Madrid, Spain.
   \email{javier.r.goicoechea@csic.es}
         \and
Institut de Radioastronomie Millim\'etrique (IRAM), 300 rue de la Piscine, F-38406 Saint Martin d’H\`eres, France.
         \and
Sorbonne Universit\'e, Observatoire de Paris, Universit\'e PSL, \'Ecole Normale Sup\'erieure, CNRS, LERMA, F-75014, Paris, France.
        \and
OASU/LAB-UMR5804, CNRS, Université Bordeaux, allée Geoffroy Saint-Hilaire, 33615 Pessac, France.
        \and
Department of Astronomy, University of Maryland, College Park, MD 20742, USA.        
 }        

   \date{Received June 4, 2018\,/\, Accepted July 28, 2018}

% \abstract{}{}{}{}{} 
% 5 {} token are mandatory    Cambiar UK to US english todo a la vez !!
 
  \abstract{We present 1$''$-resolution ALMA observations  of the circumnuclear disk (CND) and the interstellar environment around Sgr~A$^*$. The images unveil the presence of small spatial scale $^{12}$CO~($J$=3-2) molecular ``cloudlets'' ($\lesssim$20,000\,AU size) within the central parsec of the Milky Way, in other words, inside the cavity of the  CND, and moving at 
high speeds, up to \mbox{300 km\,s$^{-1}$} along the line-of-sight. 
The $^{12}$CO-emitting structures show intricate morphologies: extended and filamentary at high negative-velocities (\mbox{v$_{\rm LSR}$\,$\lesssim$$-$150~\,km\,s$^{-1}$}),  more localized and clumpy at  extreme positive-velocities 
(\mbox{v$_{\rm LSR}$\,$\gtrsim$$+$200~\,km\,s$^{-1}$}).
Based on the pencil-beam $^{12}$CO absorption spectrum toward Sgr\,A$^*$ \mbox{synchrotron} emission, we  also present evidence for a diffuse molecular gas component producing absorption features at more 
extreme negative-velocities (\mbox{v$_{\rm LSR}$\,$<$$-$200~\,km\,s$^{-1}$}).
The CND shows a clumpy spatial distribution traced by the optically thin  \mbox{H$^{13}$CN ($J$=4-3)} emission.
Its motion requires a bundle of non-uniformly rotating streams of slightly different inclinations. The inferred gas density peaks, molecular cores of a few 10$^5$~cm$^{-3}$, are  lower than the local Roche limit. This supports  that   CND cores are transient.
We apply  the two standard  orbit models, \mbox{spirals vs. ellipses}, invoked to explain the kinematics of the ionized gas streamers  around Sgr\,A$^*$. The location and velocities of the  $^{12}$CO  cloudlets inside the cavity are inconsistent with the spiral model, and only two of them 
are consistent  with the Keplerian ellipse model. Most  cloudlets, however, show  similar velocities that are incompatible with the motions of the ionized streamers or with gas bounded to the central gravity. 
We speculate that they are
leftovers of more massive molecular clouds that fall into the cavity and are  tidally disrupted, or that they  originate from instabilities in the inner rim of the CND that lead to fragmentation and infall from there. In either case, we show that molecular cloudlets, all together with a mass of several 10~$M_{\odot}$, exist around Sgr~A$^*$. Most of them must be short-lived, $\lesssim$10$^4$~yr: photoevaporated by the intense stellar radiation field, $G_0$$\simeq$10$^{5.3}$ to 10$^{4.3}$, blown away by winds from massive stars in the central cluster,   or disrupted by strong  gravitational shears.}


   \maketitle
%
%-------------------------------------------------------------------

\section{Introduction}\label{sec-introduction}

The largest baryonic--mass reservoir feeding galaxy nuclei is in the form of molecular gas. This reservoir is typically arranged as a circumnuclear disk (CND) or ring   that rotates around the \mbox{dynamical} center of the galaxy
\citep[e.g.,][and references therein]{Genzel98,Burillo14}.
   Associated with the radio-source Sgr~A$^*$, the center of the Milky Way \citep[e.g.,][]{Morris96,Genzel10} hosts a supermassive black hole (SMBH) with a mass of about 4$\cdot$10$^6$\,M$_{\odot}$ 
     \citep{Schodel02,Ghez08,Gillessen09}. It also contains a dense cluster with more than 100 OB-type massive stars  of uncertain origin \citep{Krabbe95,Schodel07,Schodel14}.
     
The gravitational potential of the SMBH is thought to influence the dynamics of a region out to half a parsec (pc) from  Sgr~A$^*$ \citep[e.g.,][]{Trani18}. The central cavity refers to the innermost pc of the Galaxy, hosting the SMBH, the nuclear stellar cluster and prominent streamers of interstellar ionized gas. The inner edge of the CND lies at about 1.5~pc from Sgr~A$^*$.     
Any material orbiting inside the cavity and close to Sgr~A$^*$ is expected to move at high velocities.
   The circular velocity in a spherically symmetric mass distribution is given by \mbox{$[G\,M(r)/r]^{1/2}$}, where $r$ is the distance to the dynamical center. In this simple view, a particle following a circular orbit would acquire velocities above 150~(300)~km\,s$^{-1}$ at  distances less than 1.0~(0.2)~pc from Sgr~A$^*$.
     
Due to a low accretion rate onto the SMBH,  the central cavity is presently quiet in terms of luminosity \citep[e.g.,][]{Baganoff03,Marrone07} and does not show evidences of a strong nuclear outflow as observed in other active galaxies 	\mbox{\citep[e.g.,][]{Cicone14}}. 
The interstellar medium of the cavity is pervaded by strong ultraviolet radiation (UV) fields and  powerful  winds from  massive  stars  \citep[][]{Morris96,Genzel10,Schodel14}. 
The combined action of these winds is thought to launch a
 roughly spherical, but currently weak, \mbox{outflow} of low-density ionized gas that interacts with the CND's inner rim \citep[][]{Geballe87,Blank16}.

\begin{figure}[t]
\centering
\includegraphics[scale=0.45, angle=0]{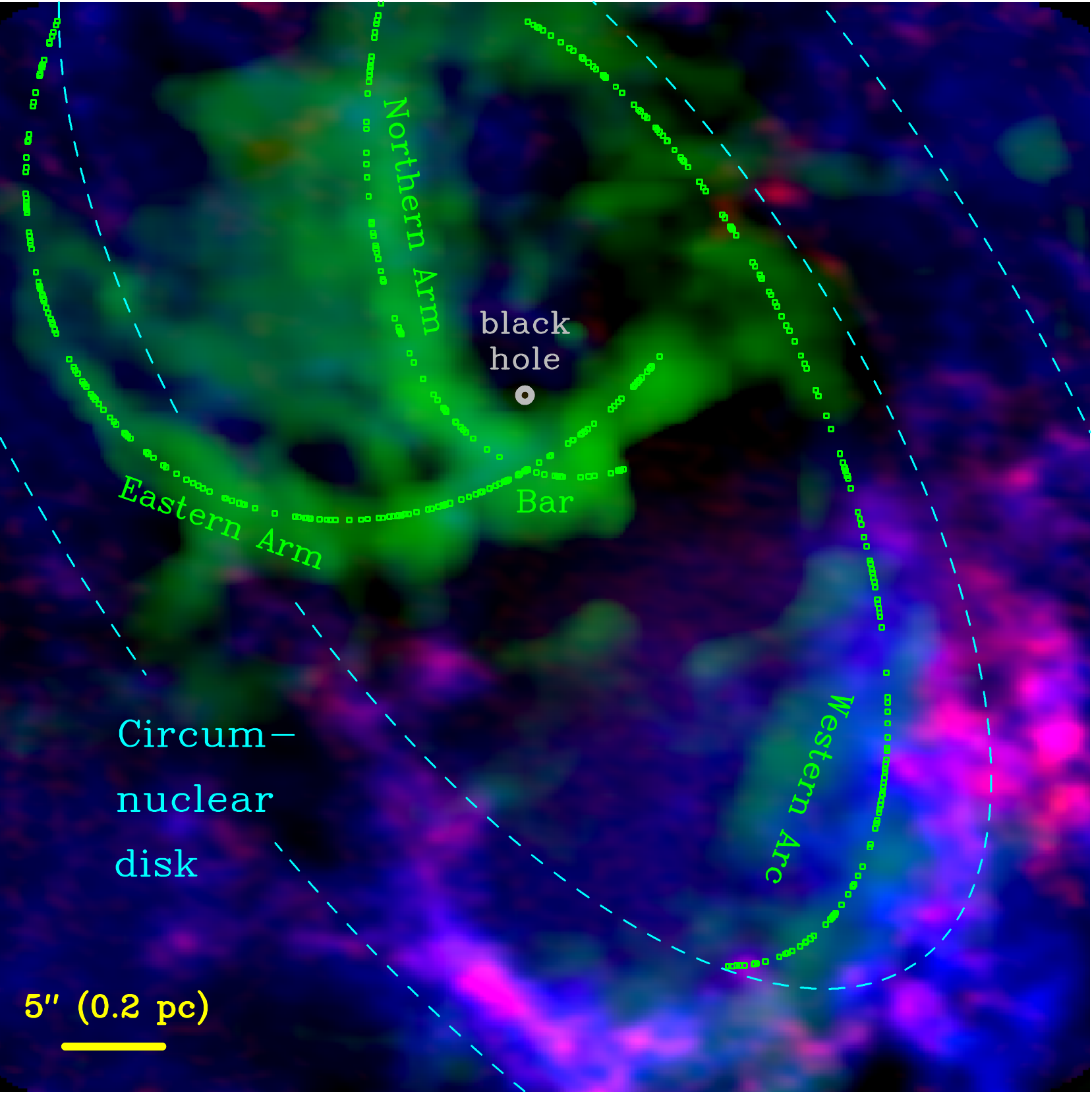}\\
\caption{Multiphase view of the central 2~pc of the Milky Way.    Overlay of the 
$^{12}$CO ($J$=3-2) (blue) and H$^{13}$CN ($J$=4-3) (red) integrated emission observed with ALMA tracing warm molecular gas in the southern lobe of the circumnuclear disk. Representative inner and outer circular orbits (1.5 and 2.3~pc deprojected radii) are shown as dashed cyan curves. Closer to the supermassive black hole at Sgr\,A$^*$, the dotted green curves show arms of the Keplerian elliptical orbits previously proposed to fit the motion of the ionized gas streamers \citep[][and references therein]{Zhao09}  shown by the [\NeII]12.8\,$\mu$m emission \citep[green, from][]{Irons12}. North is up and east is left.}\label{fig:rgb}
\end{figure}

The high-density ionized gas in the cavity is organized along three conspicuous streamers, the mini-spiral, called \mbox{western arc}, eastern arm and \mbox{northern arm}  
(see Figure~\ref{fig:rgb}), that flow around and likely  toward Sgr~A$^*$
\citep[e.g.,][]{Lacy80,Lacy91,Roberts96,Paumard04,Zhao09,Zhao10,Irons12}.  Two interpretations for the kinematics of the ionized  gas streamers have been proposed:
  
$i)$~The spiral model: nearly circular orbits with a small inward radial velocity component  leading to gas moving across the  western and northern arms \citep[e.g.,][]{Serabyn85,Lacy91,Irons12}. In addition,
\citet{Listz03} also modelled the kinematics of the ionized gas in the eastern arm and Bar region as rotating  circular rings seen nearly edge-on. 

$ii)$ The more accepted
\mbox{ellipse model}:  three families of elliptical orbits (with Sgr\,A$^{*}$ at their foci but all with different inclinations and orbital parameters) resulting in  gas flowing along the streamers. Thus, this model is  also called the tidally stretched cloud model \citep[][and references therein]{Zhao09}.

Given the observed intricate velocity patterns and complexity of the region, however,  none of these kinematic models perfectly reproduce the observed velocities. Some of the discrepancies may be related to dynamical effects produced by the presence of strong magnetic fields \citep[e.g.][]{Morris96}.  	

     Enclosing the central cavity is a rotating ring, the CND, of warm molecular gas and dust: dashed cyan curves in \mbox{Figure~\ref{fig:rgb}}, with deprojected inner and outer radii of about 1.5 and 2.5~pc respectively, inclined by $\simeq$70$^{\circ}$ and tilted on the sky by $\simeq$30$^{\circ}$ \citep{Gusten87,Jackson93,Christopher05,Montero09,Requena12,Lau13,Tsuboi18}. 
 A lower excitation molecular gas component seems to extend the CND to much larger distances, out to about 10~pc, where molecular clouds there can be tidally captured, and be dragged toward the inner ring region \citep[see e.g.,][]{Oka11,Hsieh17,Trani18}.

While many observations of the gas and dust toward the CND exist \citep[e.g.,][and references therein]{Ferriere12} it has been traditionally questioned whether molecular gas, the fuel to form new stars, can exist inside the cavity.
The existence of molecular material in such a hostile environment: affected by enhanced UV fields, X-rays,  cosmic-rays particles, stellar winds, shocks, turbulence and strong gravity effects, was initially not expected. However, it has been more or less convincingly demonstrated  recently \citep[e.g.,][]{Yusef01,Herrnstein02,Goicoechea13,Goicoechea18,Goto14,Feldmeier15,Moultaka15,Ciurlo16,Moser17,Mills17,Yusef17}.
Still,  controversy prevails because it is not easy to assign unprojected distances from the observed molecular features to Sgr\,A$^*$. The nature of the observed molecular gas emission is neither evident, interstellar: clouds or blobs, versus circumstellar: envelopes around evolved stars or young protostellar outflows and  their disks \citep[][and references therein]{Yusef17}. 
High-angular resolution, velocity-resolved  observations  with ALMA provide new avenues to probe and characterize the molecular gas component in the vicinity of a SMBH.

Using early ALMA  observations, \citet{Moser17} presented the serendipitous detection of molecular line emission (CS, SiO, C$_2$H, SO, and H$^{13}$CO$^+$) toward the inner 20$''$ of the Galaxy.
While most of the molecular line luminosity appears in the velocity range
\mbox{v$_{\rm LSR}$$\approx$$-$100 to $+$100~km\,s$^{-1}$} typical of the CND, a few CS ($J$=5-4) and SiO ($J$=8-7) emission clumps emit at significantly higher velocities: up to $\simeq$200~km\,s$^{-1}$. 
Two of these clumps were previously reported in  SiO ($J$=5-4) by \citet{Yusef13}. The
later authors interpreted the emission as arising in molecular outflows from embedded massive protostars. This interpretation is controversial because it implies \mbox{on-going} massive star formation close to Sgr~A$^*$.  Instead, \citet{Moser17} suggested that the observed features might be the dense leftovers of a disrupted molecular clump.
The presence of \mbox{bipolar outflows} and disks around low-mass protostars  within the central pc of Sgr~A$^*$ has  been also invoked from a recent analysis of ALMA 
\mbox{$^{13}$CO ($J$=2-1)} emission features, all with  $|$v$_{\rm LSR}|$$<$200~km\,s$^{-1}$ \citep{Yusef17}. However, higher angular resolution and more molecular diagnostics are needed to confirm the nature of these features. In addition, previous interferometric observations were not sensitive
to the presence of extended gas emission in the field 
(observations did not include zero- and short-spacing data).

%--------------------------------------------------------------------------------
%--------------------------------------------------------------------------------  

In this work  we present new ALMA \mbox{$^{12}$CO~($J$=3-2)} emission images of the region, complemented with \mbox{H$^{13}$CN~($J$=4-3)}
and \mbox{SO~(8$_9$-7$_8$)} images, as well as with $^{12}$CO~($J$=3-2) absorption spectra toward the strong continuum source at Sgr~\,A$^*$. 
%Because $^{12}$CO is the most abundant molecule
%after H$_2$, these observations trace much lower column densities of low-excitation molecular gas than previous studies. 
The images unveil a collection of emission features
at more extreme velocities than reported before.
Here we use the term \mbox{\textit{extreme}} \mbox{($|$v$_{\rm LSR}$$|$$\gtrsim$200~km\,s$^{-1}$)} in the context of the molecular gas emission or absorption. In particular, we argue that by virtue of its extreme velocity, such gas must be located inside the cavity and close to Sgr~A$^*$.
We note, however, that the emission from \mbox{ionized gas} inside the cavity  covers an even higher velocity range: \mbox{$-400$\,$<$v$_{\rm LSR}$$<$$+$400~km\,s$^{-1}$}  \citep[e.g.,][]{Zhao09}. As one goes closer
to the SMBH, many objects including stars and 
small putative clouds of ionized gas in the inner $\simeq$1$''$ 
($\simeq$0.04\,pc) are known to  move much faster, above 1000~km\,s$^{-1}$ in the line-of-sight \mbox{\citep[e.g.,][]{Gillessen09,Gillessen12}}.

%--------------------------------------------------------------------------------
\begin{figure*}[t]
\centering
\includegraphics[scale=0.53,angle=0]{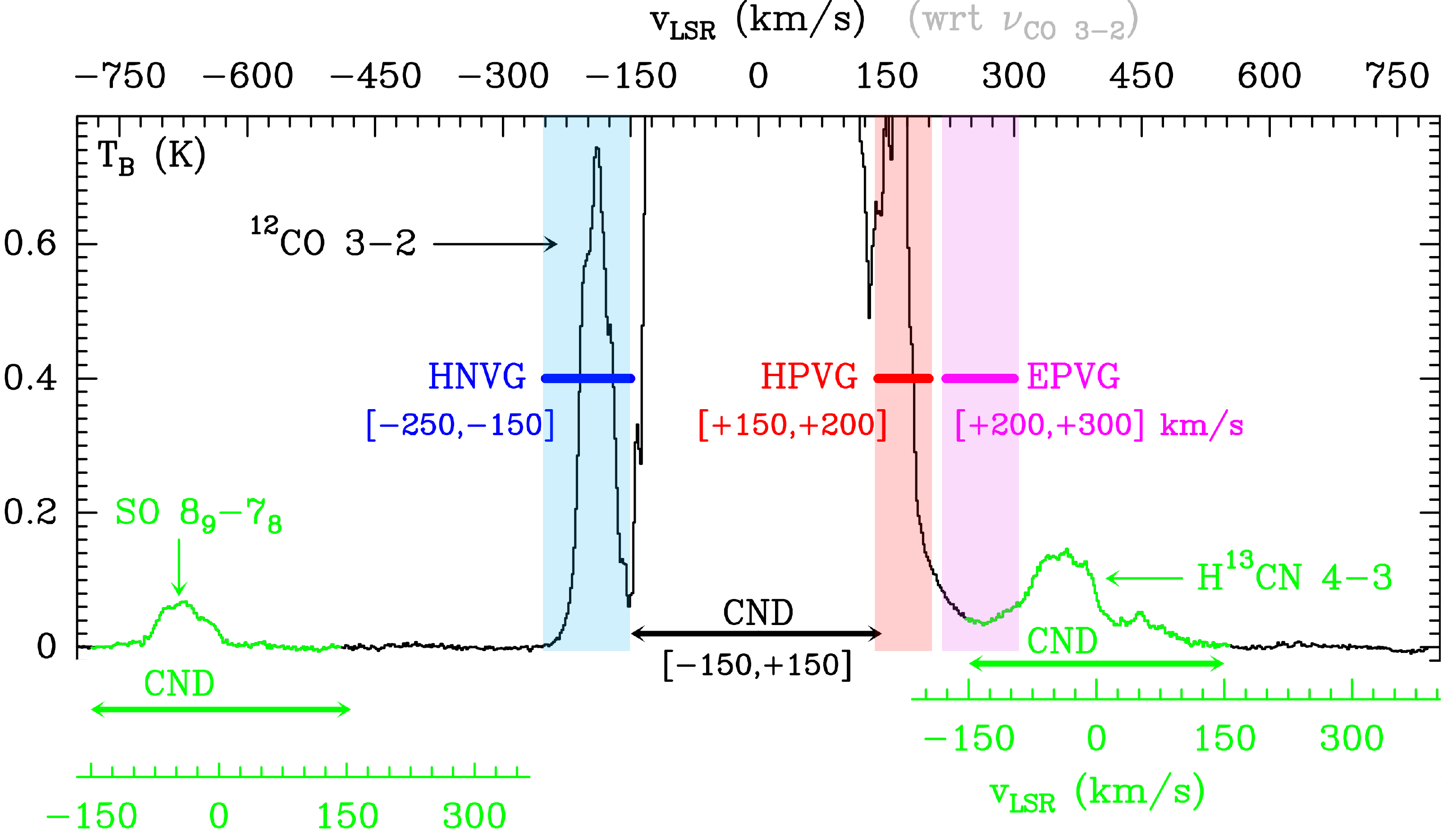}\\
\caption{ALMA plus total-power spectrum around 345.8~GHz averaged over the observed field-of-view. The black histogram  shows the spectrum centered at the frequency
of the CO~$J$=3-2 line (LSR velocity scale shown in the upper horizontal axis). The green spectra (and lower velocity axes) show the same for the \mbox{SO~8$_9$-7$_8$} and \mbox{H$^{13}$CN $J$=4-3} lines, both dominantly emitted from the CND in the velocity range \mbox{$|$v$_{\rm LSR}$$|$$<$150\,km\,s$^{-1}$} (blanked out for CO). In addition to this standard velocity range, CO shows higher positive and negative velocity
emission components: EPVG, HPVG and HNVG (defined in the text) in magenta, red, and blue respectively. Owing to the similar spatial distribution of the SO and H$^{13}$CN line emission, the CO EPVG features can be easily distinguished from the blue-shifted \mbox{H$^{13}$CN $J$=4-3} emission in the images (see text).}
\label{fig:spectrum-components}
\end{figure*}
%--------------------------------------------------------------------------------

 This paper is organized as follows. In Sect.~\ref{sec-observations} we present the ALMA data set, calibration, and image synthesis. In Sect.~\ref{sec-results}  we summarize the main observational findings. In Sec.~\ref{sec-analysis} we analyze the kinematics of the molecular gas in the CND and toward the cavity, as well as their implied masses, associated extinction and timescales for gas photoevaporation. In Sec.~\ref{sec-discussion} we discuss the possible origin of the molecular gas inside the cavity and in 
Sect.~\ref{sec-summary} we conclude.

%--------------------------------------------------------------------------------
\begin{figure*}[t]
\centering
\includegraphics[scale=0.225,angle=0]{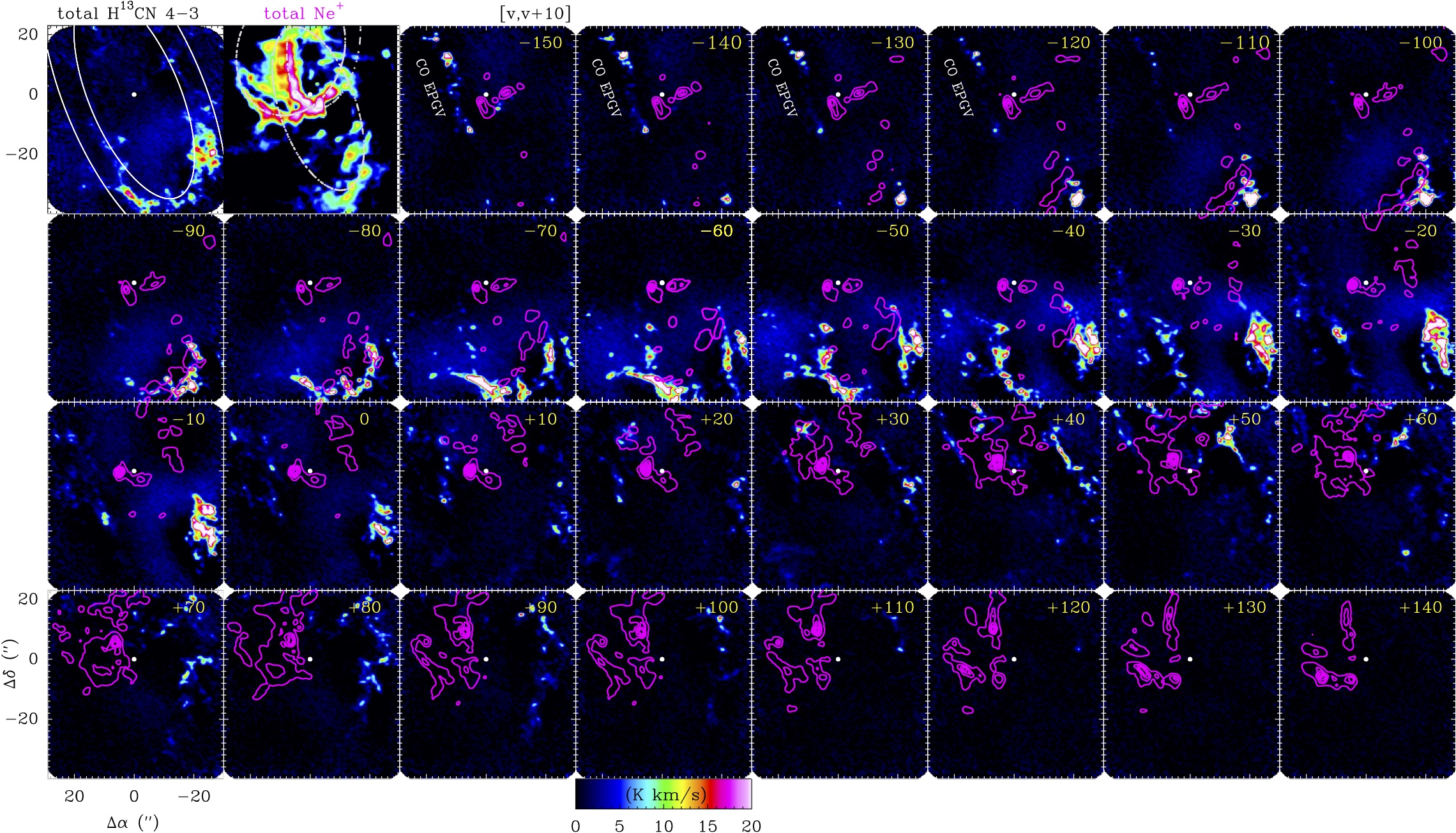}\\
\caption{ALMA H$^{13}$CN~($J$=4-3) and IRTF [\NeII]12.8\,$\mu$m velocity channel  maps 
from v$_{\rm LSR}$=$-$150 to $+$150~km\,s$^{-1}$ in steps of 10~km\,s$^{-1}$. 
 The two top left panels show H$^{13}$CN~($J$=4-3)
and [\NeII]12.8\,$\mu$m total integrated line intensity maps.
 White curves in the H$^{13}$CN~($J$=4-3) map represent a CND model bounded by two rotating circular rings (with $\iota$=66$^{\circ}$, $\theta$=23$^{\circ}$) with inner and outer deprojected radii of 1.5 and 2.3 pc respectively. Gray curves in the [\NeII]12.8\,$\mu$m
map represent Keplerian elliptical orbits proposed to fit the motion of the ionized gas streamers \citep{Zhao09}.
The other panels show individual H$^{13}$CN velocity channel maps (for intensities between 0 to 20~K\,km\,s$^{-1}$ in linear scale) with the specific velocity interval $[$v,v+10$]$ displayed at each corner. The ionized gas emission  in the mini-spiral is shown in magenta contours \citep[from][]{Irons12}. 
Sgr\,A$^*$~is at position (0$''$,0$''$) and is displayed with a white dot.
}
\label{fig:h13cn}
\end{figure*}
%--------------------------------------------------------------------------------

\section{Observations}\label{sec-observations}

Observations toward Sgr~A$^*$ (\mbox{R.A.\,(J2000)=17h\,45\,m\,40.036s};  \mbox{DEC.\,(J2000)=-29$^{\circ}$00$'$28.17$''$}) and the southern lobe of circumnuclear disk \citep[CND-S,][]{Becklin82} were carried out using 28 \mbox{12 m-array} antennas as part of project 2013.1.01058.S \mbox{(PI. J.~R. Goicoechea)}. The target line frequencies of this work are at 345.796~GHz (\mbox{CO $J$=3–2)}, 345.339~GHz (\mbox{H$^{13}$CN $J$=4-3}), and 346.528~GHz (\mbox{SO 8$_9$-7$_8$}). 
The observations consisted of a 39-pointing mosaic 
(see footprints in Figure~\ref{fig:h13cn_tp_appendix} of the \mbox{Appendix}).
The field-of-view is \mbox{$60''\times60''$}. Baseline configurations from 12 to about 350~m were measured. We employed a correlator providing a resolution of 975~kHz ($\simeq$0.8\,km\,s$^{-1}$) over an 1875~MHz bandwidth ($\simeq$1600\,km\,s$^{-1}$), that is, the three target lines appear in the same spectrum. Figure~\ref{fig:spectrum-components} shows  the frequency coverage
of the observations. The observation time with the ALMA 12 m-array was $\simeq$3~h. ALMA executing blocks were first calibrated in  CASA~4.6.0 software  and then exported to GILDAS software for imaging, deconvolution, and analysis.  
    In order to create a realistic image of the observed field and to recover any extended emission component, we added as zero-spacing a deep, fully-sampled single-dish map  taken with 3 different 12-m antennas providing a beam full-width at half-maximum (FWHM) of about 18$''$. The total on-source time of these single-dish \mbox{‘total power’} (TP) map was 17.4~h. We used the on-the-fly mode with a distant reference position located at \mbox{$\Delta$R.A.=$+$258$''$} and  \mbox{$\Delta$DEC.=$-$978$''$}. The rms noise of the single-dish map is about 7~mK per 2~km\,s$^{-1}$ channel. Figure~\ref{fig:h13cn_tp_appendix} in the Appendix shows the TP map used for H$^{13}$CN~($J$=4-3) line. Because only a single reference position was observed and the $^{12}$CO emission is widespread and detectable throughout the Galactic center, we expect that residual signal in the reference position can negatively contaminate the  $^{12}$CO emission between v$_{\rm LSR} \approx -100$ and $+$100~km\,s$^{-1}$. Thus, in this work  we only discuss the $^{12}$CO emission in the 
\mbox{$|$v$_{\rm LSR}$$|$$>$130\,km\,s$^{-1}$}  range. On the other hand, the \mbox{$J$=4-3} line of the trace species H$^{13}$CN is faint and optically thin \citep[cf.][]{Christopher05,Montero09,Mills13}, thus not affected by foreground/background contribution: it arises only from the CND and surroundings. 

%--------------------------------------------------------------------------------
\begin{figure*}[ht]
\centering
\includegraphics[scale=0.225,angle=0]{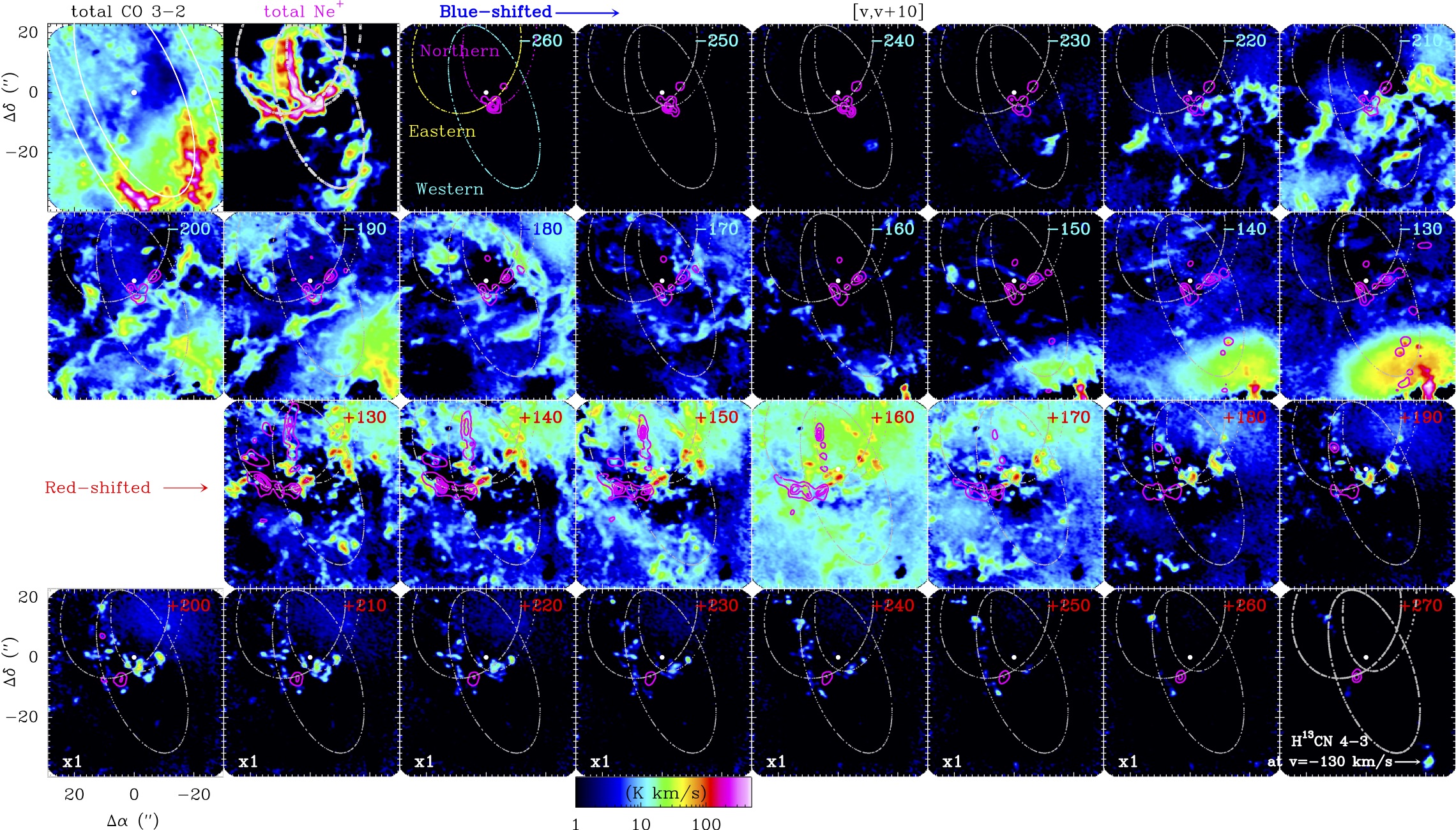}\\
\caption{ALMA high-velocity $^{12}$CO~($J$=3-2) and IRTF [\NeII]12.8\,$\mu$m high-velocity channel  maps 
from v$_{\rm LSR}$=$-$260 to $-$120~km\,s$^{-1}$ (blue-shifted) and
from v$_{\rm LSR}$=$+$130 to $+$280~km\,s$^{-1}$ (red-shifted) in steps of 10~km\,s$^{-1}$. 
 The two top left panels show  $^{12}$CO~($J$=3-2)
and [\NeII]12.8\,$\mu$m total integrated line intensity maps.
 White curves in the $^{12}$CO~($J$=3-2) map represent a CND model bounded by two rotating circular rings (with $\iota$=66$^{\circ}$, $\theta$=23$^{\circ}$) with inner and outer deprojected radii of 1.5 and 2.3 pc respectively. Gray curves represent the three Keplerian elliptical orbits  proposed to fit the motion of the ionized gas streamers \citep{Zhao09}.
All the other panels show $^{12}$CO velocity channel maps (for intensities between 1 to 500~K\,km\,s$^{-1}$ in logarithm scale) with the specific velocity interval $[$v,v+10$]$ displayed at each corner. The ionized gas emission  in the mini-spiral is shown in magenta contours \citep[from][]{Irons12}. 
Sgr\,A$^*$~is at position (0$''$,0$''$) and is displayed with a white dot.}
\label{fig:co-vchannels}
\end{figure*}
%--------------------------------------------------------------------------------

     The GILDAS/MAPPING software was used to create the zero-spacing visibilities \citep{Pety10} not sampled by the ALMA array. 
    %For the primary beam correction in the mosaicing processes we used a %Gaussian function of 16.8$''$ FWHM
%truncated at 20\%~of the peak.  
     These pseudo-visibilities were merged with the interferometric (alone) observations in the UV plane. A baseline subtraction was performed aslo in the UV plane on the \mbox{ALMA-12m} visibilities before merging. 
To do that, we created a linear velocity axis with 800 channels of 2~km~s$^{-1}$ width centered at the rest frequency of the $^{12}$CO~$J$=3-2 line, and we made a linear fit to velocity
windows devoid of molecular line emission. In particular, we 
excluded the \mbox{$[-800,-300]$} and \mbox{$[-250,+560]$~km\,s$^{-1}$} velocity ranges (i.e., avoiding the SO, $^{12}$CO, and  H$^{13}$CN line emission, see
Fig.~\ref{fig:spectrum-components}).  
     Each mosaic field was then imaged and a dirty mosaic was built. The dirty image was deconvolved using the 
H\"ogbom CLEAN algorithm. 

The resulting cubes, smoothed by integrating rectangles, to a velocity resolution of 2~km\,s$^{-1}$, were scaled from Jy/beam to brightness temperature scale using the final synthesized beam-size of 1.1$''$$\times$0.7$''$. 
The rms noise in the final cube is $\sim$200\,mK per resolution channel.
Figures~\ref{fig:h13cn_so_only_alma_appendix} and \ref{fig:h13cn_so_appendix} in the Appendix show \mbox{H$^{13}$CN $J$=4-3} and \mbox{SO 8$_9$-7$_8$} velocity channel maps obtained from the ALMA-alone and the ALMA+TP data sets 
respectively. \mbox{Figures~\ref{fig:co_alma-only_appendix}} and \ref{fig:co_alma-tp_appendix} show the high and extreme positive- and 
negative-velocity $^{12}$CO channel maps, again for the ALMA-alone and ALMA+TP data sets 
respectively.

\section{Results}\label{sec-results}

 %--------------------------------------------------------------------------------
\begin{figure}[h]
\centering  %0.72
\includegraphics[scale=0.18,angle=0]{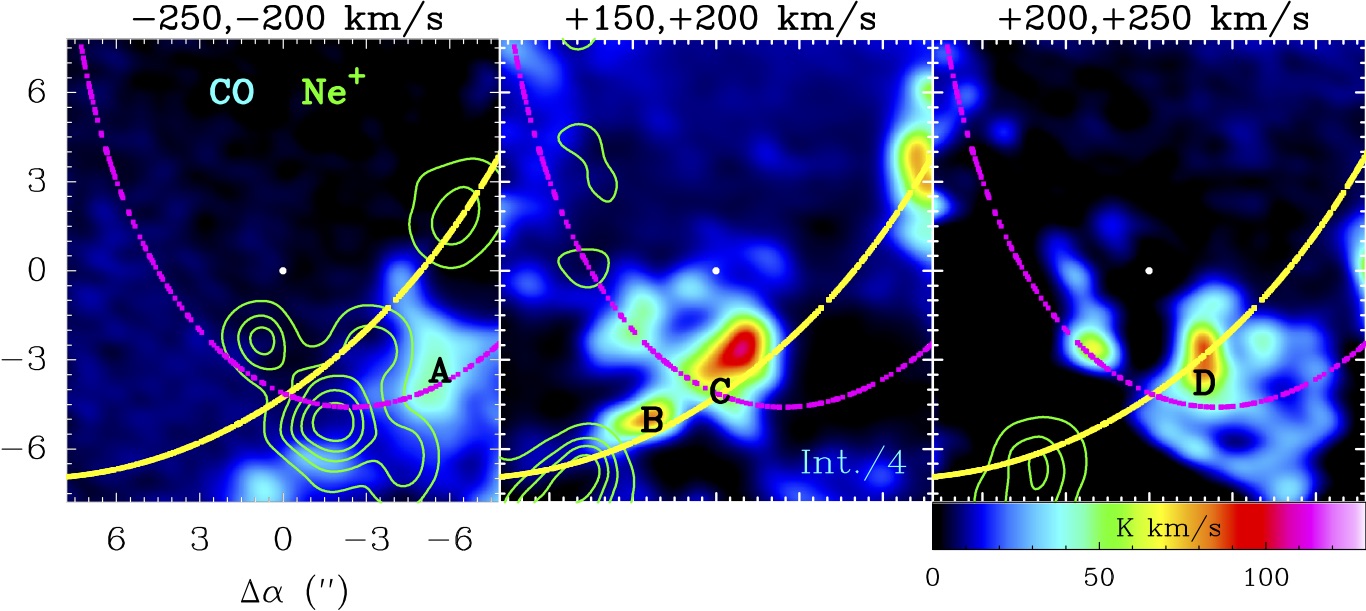}\\
\caption{Zoom into the central light-year of the Galaxy. Images show 
\mbox{CO~($J$=3-2)} high-speed channel maps (\textit{left}: v$_{\rm LSR}$=$-$250 to \mbox{-200 km\,s$^{-1}$}, \textit{middle}: v$_{\rm LSR}$=$+$150 to $+$200~km\,s$^{-1}$, and \textit{right}: v$_{\rm LSR}$=$+$200 to $+$250~km\,s$^{-1}$), showing the \mbox{integrated} line intensity from 0 to 130~K\,km\,s$^{-1}$, and revealing molecular gas  close to Sgr\,A$^*$. Green contours show the ionized gas emission
\citep[][]{Irons12}. The yellow and magenta tracks are segments of the Keplerian elliptical orbits along the eastern and northern arms respectively \citep{Zhao09}.} 
%Insets show broad-line CO cloudlet spectra (all covering a 100~km\,s$^{-1}$ bandwidth) at %representative positions.}
\label{fig:Zoom}
\end{figure}
%--------------------------------------------------------------------------------
  
 %in the range 
%\mbox{$|$v$_{\rm LSR}|$$>$200~km\,s$^{-1}$} where the $^{12}$CO emission is already faint.

% roughly the central 2~pc of the Galaxy,

Figure~\ref{fig:rgb} shows the spatial distribution of the 
\mbox{$^{12}$CO~($J$=3-2)} (blue) and \mbox{H$^{13}$CN ($J$=4-3)} (red)  integrated line intensities, arising mostly from the CND, and of the [\NeII]12.8\,$\mu$m  
line\footnote{The ionization potential of the Ne atom is 21.6~eV. The Ne$^+$ observations of \citet{Irons12} have angular and velocity resolutions of 1.3$''$ and 4~km\,s$^{-1}$ respectively, thus comparable to our ALMA images.} (green) from ionized gas 
 inside the cavity \citep[from][]{Irons12}.
\mbox{Figure~\ref{fig:spectrum-components}} shows the ALMA+TP spectrum averaged over the observed field-of-view. The velocity axes show the velocity distribution of the three detected emission lines. The green spectra show that the 
\mbox{H$^{13}$CN $J$=4-3} and \mbox{SO 8$_9$-7$_8$} emission arises
at velocities \mbox{$|$v$_{\rm LSR}|$$<$150~km\,s$^{-1}$} typical of the CND.  
Combining the spatial and velocity distributions as channel maps,  \mbox{Figure~\ref{fig:h13cn}} shows that the bulk  H$^{13}$CN ($J$=4-3) emission is dominated by clumpy structures within the southern lobe of the CND \mbox{(CND-S)}.  
The optically thin \mbox{H$^{13}$CN $J$=4-3} emission peaks trace moderately dense molecular cores %(with density peaks $n_{\rm H}$ of several 10$^5$~cm$^{-3}$}, 
(see \mbox{Sect.~\ref{subsec-analysis-masses}}).
As reported in previous  \mbox{H$^{12}$CN} observations \citep{Christopher05,Montero09}, these structures  follow a necklace distribution along the CND. Our higher angular resolution images, by more than
a factor of four,  show that the optically thick \mbox{H$^{12}$CN ($J$=4-3)} emitting structures \citep[][]{Mills13,Smith14} break into 
smaller and sharper \mbox{H$^{13}$CN ($J$=4-3)}  cores. 
In addition, there is a diffuse gas component, meaning faint and extended, spreading around the inner edge of the CND-S. The velocity channel images in Figure~\ref{fig:h13cn} visually suggest that part of this diffuse emission  are lower density tails of denser structures rotating in the CND, perhaps the leftovers of cores that have been already stretched and disrupted.

%The $^{12}$CO  emission in this velocity range is more extended (see e.g., the $-$130~km\,s$^{-1}$ channel map in Fig.~\ref{fig:co-vchannels}), tracing warmer %and lower density gas that envelops the H$^{13}$CN structures  and is still %bright in the more \mbox{UV-exposed} rim close to the ionized western arc.

But can molecular gas exist inside the CND cavity, closer to Sgr\,A$^*$, for example  within the central light-year
($\approx$0.3~pc or $\approx$7$''$) of the Galaxy?
 $^{12}$CO  is  a very  stable molecule and about 4 orders of magnitude more abundant than H$^{12}$CN. Hence,  $^{12}$CO rotational emission probes very small column densities of molecular gas, hardly detectable in any other molecule, especially at velocities \mbox{$|$v$_{\rm LSR}|$$>$150~km\,s$^{-1}$}, for which the $^{12}$CO emission is already faint. 
Indeed, the average \mbox{$^{12}$CO $J$=3-2} spectrum in Figure~\ref{fig:spectrum-components} 
shows additional red-shifted $^{12}$CO emission components that we name 
\textit{high positive-velocity gas} (HPVG; \mbox{$+150$$<$v$_{\rm LSR}$<$$+$200$~km\,s$^{-1}$}) and \textit{extreme positive-velocity gas} 
(EPVG; \mbox{$+200$$<$v$_{\rm LSR}$<$$+$300$~km\,s$^{-1}$}).
The associated \mbox{$^{12}$CO~($J$=3-2)} EPVG  frequency range coincides with the blue-shifted
\mbox{H$^{13}$CN}  emission (\mbox{v$_{\rm LSR}$(H$^{13}$CN)$\lesssim$$-$100~km\,s$^{-1}$}, see the spectrum in Fig.~\ref{fig:spectrum-components}).
Fortunately, we can easily  distinguish the  $^{12}$CO EPVG emission 
from sources of  blue-shifted \mbox{H$^{13}$CN} emission
 by taking advantage of the nearly identical spatial and velocity distribution of the H$^{13}$CN~$J$=4-3 and \mbox{SO~8$_9$-7$_8$} lines (see channel maps in Figs.~\ref{fig:h13cn_so_only_alma_appendix} and \ref{fig:h13cn_so_appendix} of the Appendix). In particular, 
the emission peaks seen in the range v$_{\rm LSR}$$\simeq$$+$200 to $+$300~km\,s$^{-1}$, with respect to  \mbox{$^{12}$CO~$J$=3-2}, but  not detected in \mbox{SO 8$_9$-7$_8$}, are real \mbox{$^{12}$CO} EPVG emission features and not \mbox{H$^{13}$CN ($J$=4-3}). On the other hand, there are also features  that do arise from 
  \mbox{blue-shifted}  
 \mbox{H$^{13}$CN $J$=4-3} emission (v$_{\rm LSR}$$\simeq$$-$100 to $-$140~km\,s$^{-1}$), specially  toward the southern lobe of the CND because they are also
 detected in \mbox{SO~8$_9$-7$_8$} emission (see  Figs.~\ref{fig:h13cn_so_only_alma_appendix} and \ref{fig:h13cn_so_appendix}).

\begin{figure*}[t]
\centering
\includegraphics[scale=0.8,angle=0]{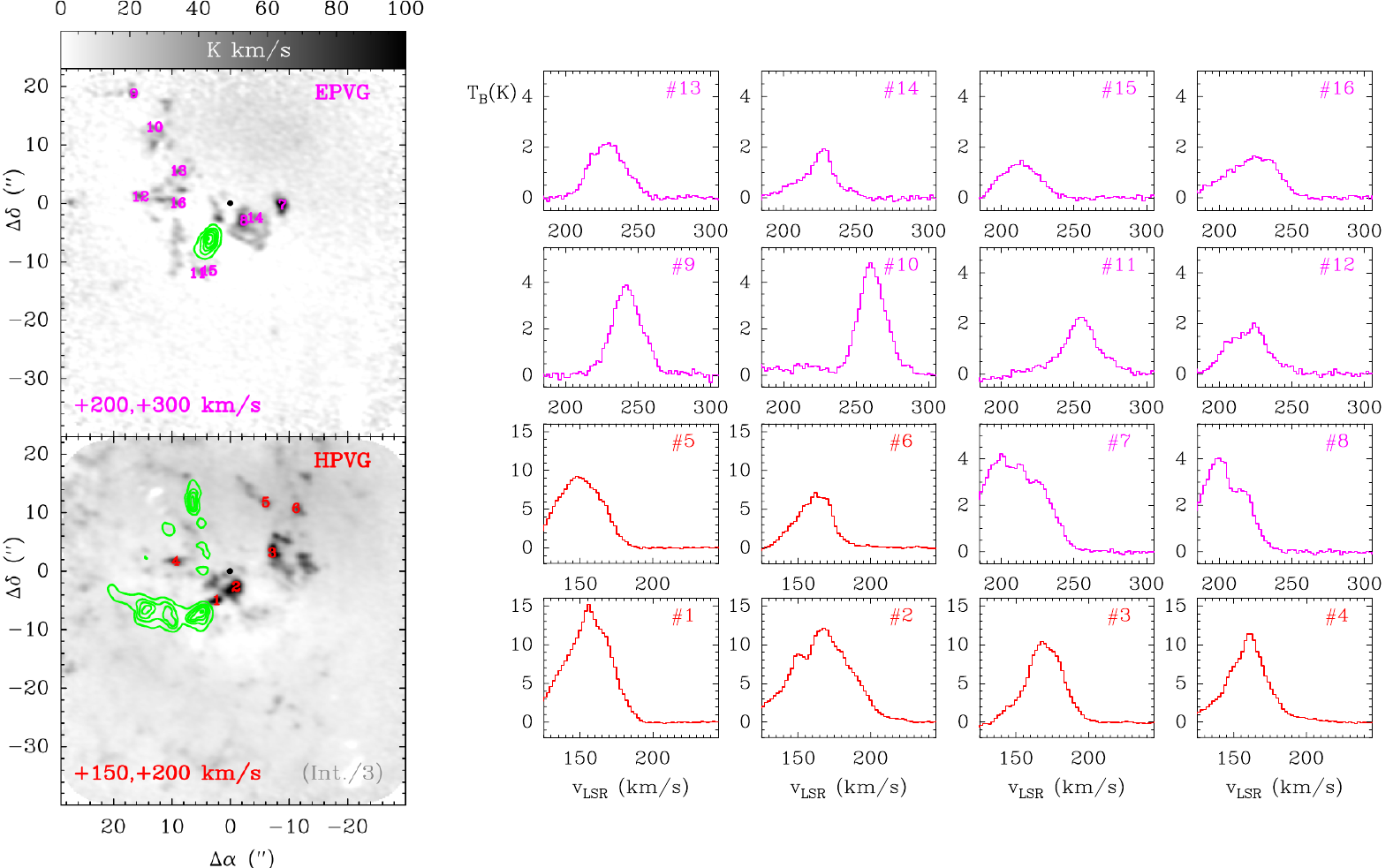}\\
\caption{\textit{Left:} EVPG (magenta, top)  and HPVG (red, bottom)  molecular cloudlets, numbered, for which we
have extracted their $^{12}$CO~$J$=3-2  spectra. The gray images show the $^{12}$CO $J$=3-2 line emission integrated in the velocity interval displayed in each panel. The green contours shows the ionized gas emission in the same velocity range ([\NeII]12.8~$\mu$m line from Irons et al. 2012). Sgr\,A$^*$~is  displayed with a black dot.
\textit{Right:}~EVPG (magenta) and HPVG (red)  cloudlet spectra. The numbers correspond to their position in the left panels. In each panel the velocity axis has a common width
of 125~km\,s$^{-1}$. Gaussian line fit parameters are tabulated in 
Table~\ref{table:cloudlets}.}
\label{fig:cloudletfinder}
\end{figure*}
%--------------------------------------------------------------------------------

Figures~\ref{fig:co-vchannels} and \ref{fig:Zoom} %%and \ref{fig:cloudletfinder}
show the collection of  HPVG and EPVG clumpy emission structures detected toward the cavity interior.
\mbox{Owing to} their small sizes ($\approx$2$''$$\lesssim$20,000\,AU), we term these structures molecular \textit{cloudlets}.  Their (projected) location on the sky   apparently agrees with the elliptical orbits  of the ionized streamers inside the cavity (Fig.~\ref{fig:co-vchannels}): $i)$~the~brightest cloudlets in the range \mbox{$+$100$<$v$_{\rm LSR}$<$ +200$~km\,s$^{-1}$} resemble an extension of the eastern arm going northwest upstream,  $ii)$ at more extreme velocities, many EPVG cloudlets are distributed northeast of Sgr\,A$^*$, apparently along the western orbit, and
$iii)$ close to the Bar, the  $^{12}$CO emission  shows a bundle of emission features (structure D in Fig.~\ref{fig:Zoom}). %reminiscent of a cloudlet-cloudlet collision. 
In the next section we compare the velocities of these  cloudlets with the kinematics of the ionized gas to determine whether these  spatial coincidences are fortuitous or both the molecular  and the ionized streamers are linked.

In Figure~\ref{fig:cloudletfinder} we identify the location and extract the spectra  of several  $^{12}$CO HPVG and EPVG cloudlets: emission peaks that show an isolated spectral component in a given velocity range; Gaussian line fit parameters are tabulated in  \mbox{Table~\ref{table:cloudlets}} of the Appendix. Despite their small sizes, their line-profiles are very broad,
typically \mbox{$\Delta {\rm v}$$\simeq$25-40~km\,s$^{-1}$}, suggesting 
 strong tidal shear and turbulence dissipation. We checked that 
their position and velocities do not coincide with the \mbox{proplyd} and
low-mass protostar outflow candidates reported by \mbox{\citet{Yusef17b,Yusef17}} inside the cavity.  
Only the $^{12}$CO HPVG cloudlet~$\#$4 coincides  with the SiO emission ``Clump~2'' previously
detected by \citet{Yusef13} and \citet{Moser17} southwest of IRS\,1W star cluster.
  
Although we added the TP data to recover the extended molecular emission filtered out by the interferometer, the  bulk  of the $^{12}$CO
clumpy emission in the EPGV range comes from small-scale structures detected by  ALMA interferometer alone. This emphasizes their compact sizes  
(Fig.~\ref{fig:co_alma-only_appendix}). In general, the TP data adds a more extended and faint gas component that surrounds the H$^{13}$CN cores in the CND, and also the $^{12}$CO cloudlets  at higher velocities. In the Appendix we show TP and \mbox{ALMA-alone} velocity channel maps.

In addition to the positive-velocity emission components, we also detect widespread $^{12}$CO emission at high negative-velocities  
(v$_{\rm LSR}$$<$$-$150~km\,s$^{-1}$, see Figure~\ref{fig:co-vchannels}). These structures are more  extended and show very filamentary morphologies.
Their exact location, their distance to Sgr~A$^*$, is much more
difficult to determine because this is also the velocity range of the so-called 
foreground high negative-velocity gas 
(HNVG; \mbox{$-$200$\lesssim$v$_{\rm LSR}$$\lesssim$$-$150~km\,s$^{-1}$}) clouds \citep[e.g.,][]{Gusten81,Zhao95}.  These foreground clouds are not related to the CND or Sgr~A$^*$ (see \mbox{Section~\ref{subsec-analysis-absorption}}).
Nevertheless,   some of the observed  blue-shifted $^{12}$CO cloudlets seem to lie inside the cavity because they overlap or closely delineate the shape of some ionized gas features, for example structure A in 
Fig.~\ref{fig:Zoom} near the Bar region where the northern and eastern arms intersect.

%\clearpage

\section{Analysis}\label{sec-analysis}

%--------------------------------------------------------------------------------
\begin{figure}[h]
\centering
\includegraphics[scale=0.43,angle=0]{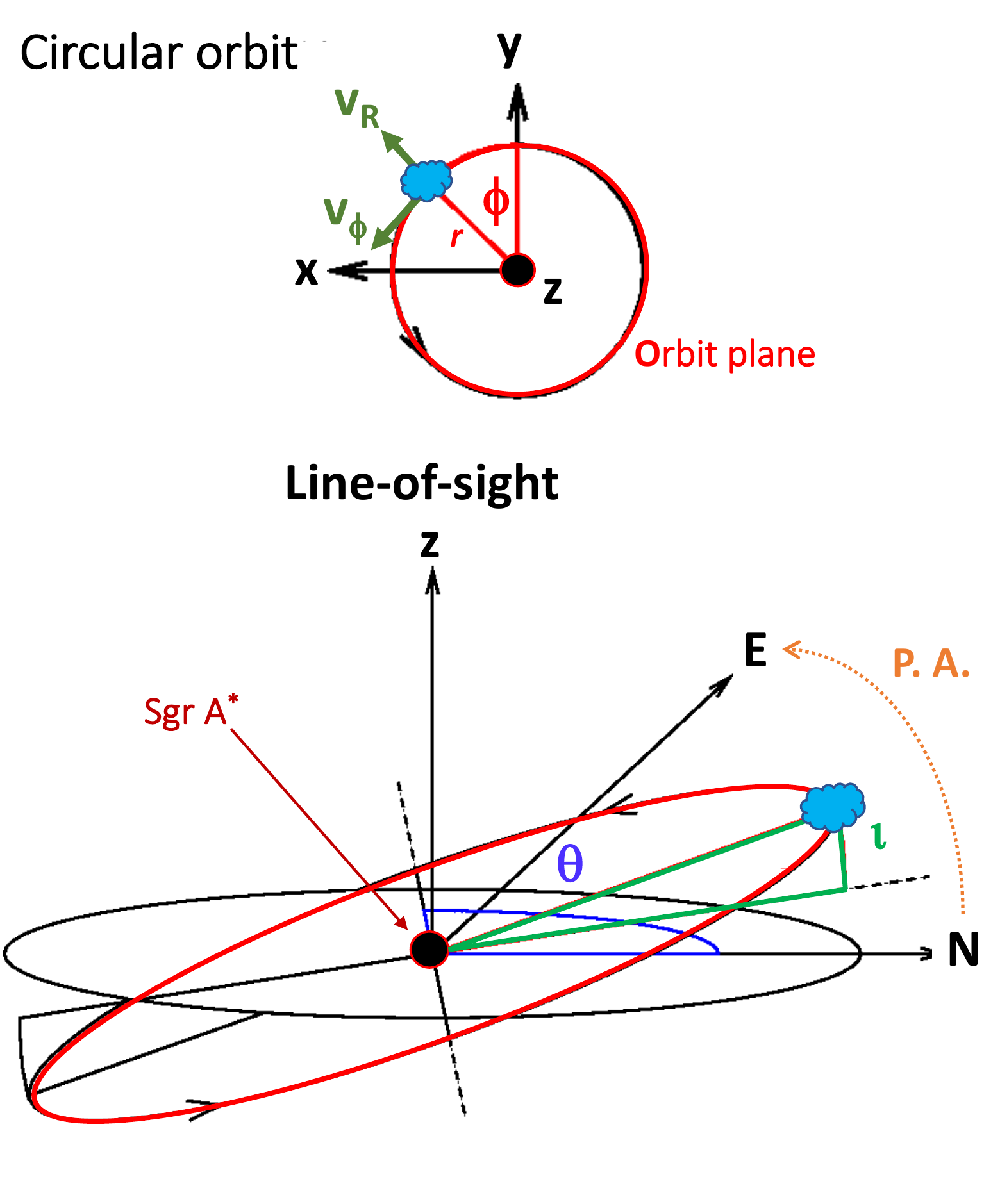}\\
\caption{Cloudlet in a circular orbit. \textit{Orbit plane:} The position of
a cloudlet is determined by the azimuth angle $\phi$ and the deprojected distance
$r$ to Sgr~A$^*$. The circular (tangential) velocity is v$_{\phi}$. A small
radial velocity v$_{\rm R}$ will make the cloudlet spiral outward (v$_{\rm R}$$>$0) or inward (v$_{\rm R}$$<$0). \mbox{\textit{Plane of the sky:}} The  orbit appears to an observer as an
ellipse characterized  by the inclination angle
$\iota$ between the two planes, and by the angle $\theta$ between the north axis and the intersection of the orbital and sky planes. P.\,A. refers to the position angle on the sky, from the north axis to the east axis, used in Figure~\ref{fig:pos_angle_velocities} and in the text.}
\label{fig:circular_orbit}
\end{figure}
%--------------------------------------------------------------------------------

\subsection{CND kinematics: bundle of non-uniformly rotating orbits}\label{subsec-discussion-analysis}

%--------------------------------------------------------------------------------
\begin{figure*}[h]
\centering
\includegraphics[scale=0.22,angle=0]{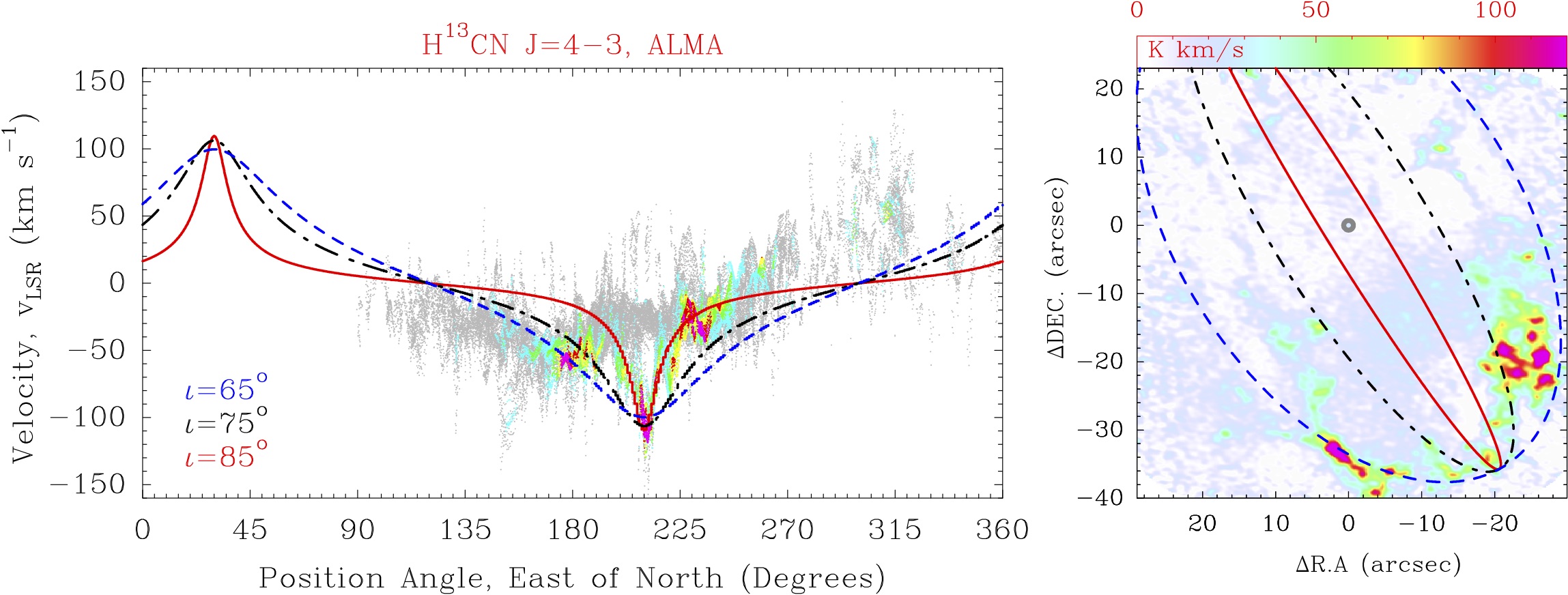}\\
\caption{Mean H$^{13}$CN ($J$=4-3) velocities and models of rotating circular rings for the CND. \textit{Left}: observed velocities as a function of position angle on the sky (from north to east). Colors represent the same intensity scale as in the integrated line intensity image shown in the right panel. Curves show models of rotating, tilted rings with a circular velocity v$_{\phi}$=110~km\,s$^{-1}$, $\theta$=30$^{\circ}$, and different inclination angles $\iota$. 
\textit{Right}:~Integrated intensity map and resulting orbits (with a deprojected radius of 1.7~pc in this example).}
\label{fig:pos_angle_velocities}
\end{figure*}
%--------------------------------------------------------------------------------

As common in the literature, we model the CND kinematics as a set of uniformly rotating rings \citep{Gusten87,Jackson93}.  We adopt a distance of 8300~pc to Sgr\,A$^*$ \mbox{\citep{Genzel10}}.
The sketch in \mbox{Figure~\ref{fig:circular_orbit}} summarizes the coordinate system.
 Assuming circular velocities, with $\phi$ the \mbox{azimuth} angle in the orbit plane, the projected coordinates on the sky and those of the orbital plane are related by the inclination angle $\iota$ between the two planes, and by the angle $\theta$ between the north axis and the intersection of the orbital and sky planes. 
 The observed line-of-sight velocity (v$_{\rm LSR}$) at a given position of the projected orbit is given by 
\mbox{v$_{\rm LSR}$($r$)=v$_{\phi}(r)\,sin\,\iota \,cos\,\phi$}, 
where v$_{\phi}$ is the circular (tangential) velocity in the orbital plane \citep{Rogstad74} and where we take into account that the source systemic velocity is \mbox{v$_{\rm SYS}$=0~km\,s$^{-1}$}. The circular velocity of a particle moving in a gravitational potential is determined by the enclosed mass at a given radius $r$ from the \mbox{dynamical center}, with 
\mbox{v$_{\phi}=[G\,M(r)/r]^{1/2}$}
and \mbox{$M(r)=M_{\rm SMBH}+M_{\star}(r)$}. This expression accounts for the mass of the SMBH at Sgr\,A$^*$ \citep[\mbox{$M_{\rm SMBH}$$\simeq$4.2$\cdot$10$^6$\,M$_\odot$},][]{Schodel02,Ghez08,Gillessen09} and for the accumulated mass of the nuclear stellar cluster  at a given radial distance from the center $M_{\star}(r)$ \citep[e.g.,][]{Genzel10,Schodel14}. Here we follow \citet{Irons12} and use a simplified mass distribution \mbox{$M_{\star}(r) \simeq m_{\star}\cdot r^{1.2}$} with
 $r$ in pc and \mbox{$m_{\star}=10^6\,M_{\odot}$} \citep[][]{Genzel10}. In order to improve the fit, we allowed an ad hoc parameter $\beta$, \mbox{from 0 to 1}, so that the actual \mbox{circular velocity is v$_{\phi} \cdot \beta$}.  For $\beta <1$, this is equivalent to decreasing the enclosed mass but it can also be interpreted as a brake mechanism due to friction, perhaps as a result of  enhanced magnetic fields in the CND. %\citep[e.g.,][]{Morris96}.

Figure~\ref{fig:pos_angle_velocities}  shows synthetic rotation curves plotted over the 
H$^{13}$CN~($J$=4-3) intensity-weighted mean velocities as a function of position angle on the sky (P.A.~from north to east, see sketch in \mbox{Fig.~\ref{fig:circular_orbit}}).  Most of the bright H$^{13}$CN~($J$=4-3) emission can be bracketed by models of circular orbits tilted by $\theta$$\simeq$30$^{\circ}$ and $\iota$ from 65$^{\circ}$ to 75$^{\circ}$ but 
with indications of  orbits of higher inclination,  
$\iota$$\simeq$85$^{\circ}$. Our inferred 
$\iota$ and $\theta$ angles  are slightly larger than previous estimations
\citep[e.g.,][]{Gusten87,Jackson93,Smith14} and show that, even for the typical 
CND bright emission velocities \mbox{($|$v$_{\rm LSR}| \lesssim 100$~km\,s$^{-1}$)} the gas motion is better described by a bundle of orbits that cover different inclinations instead of a co-planar set of circular orbits.  
In agreement with previous low-resolution observations \citep{Gusten87,Jackson93}, the required circular velocity is \mbox{v$_{\phi}$=100-110~km\,s$^{-1}$} ($\beta$$\simeq$0.8-0.9). In the inner edge of the CND, this circular motion implies a rotational period $\tau_{\rm rot}(\varphi)$ of $\lesssim$10$^5$~yr. 
\mbox{Figure~\ref{fig:pos_angle_velocities}} also demonstrate that a single rotating structure can’t fit the H$^{13}$CN emission from the CND \citep[see also][]{Martin12}. 

%From the observations and model comparison 
It is also clear that several structures in the CND depart from a uniformly rotating circular motion.
In particular, the emission structure emitting
at \mbox{v$_{\rm LSR}$$\simeq$$+$40 to $+$100~km\,s$^{-1}$} (\mbox{P.A.$\simeq$300-315$^{\circ}$} in Figures~\ref{fig:h13cn} and \ref{fig:pos_angle_velocities}). This component may
be related to the $+$80~km\,s$^{-1}$ OH-absorption streamer detected by
\mbox{\citet{Karlsson15}} and that likely follows a specific trajectory. 
In addition,  Figure~\ref{fig:pos_angle_velocities} shows the contribution of the diffuse H$^{13}$CN emission component that spreads around the inner edge of the CND-S (gray color shaded areas in Fig.~\ref{fig:pos_angle_velocities}).
Most  positions and velocities associated to this  component cannot be fitted by circular orbits either.

With the [\NeII]12.8\,$\mu$m cube in hand, we also analyze the kinematics of the ionized gas adjacent to the CND-S, that is, the~western arc.
Globally, the ionized and the molecular gas  follow the same velocity pattern
(P.A. from $\simeq$170$^{\circ}$ to 230$^{\circ}$ in Fig.~\ref{fig:pos_angle_velocities}). In particular, the channel maps
v$_{\rm LSR}$$\approx$$-$100 to $-$50~km\,s$^{-1}$
in Figure~\ref{fig:h13cn}
show the smooth transition between the ionized gas and the H$^{13}$CN emission, confirming previous suggestions that the western arc is the inner
ionized  rim of the \mbox{UV-irradiated}  CND-S lobe \citep{Zhao09,Irons12}.

%--------------------------------------------------------------------------------
\begin{figure*}[t]
\centering
\includegraphics[scale=0.18,angle=0]{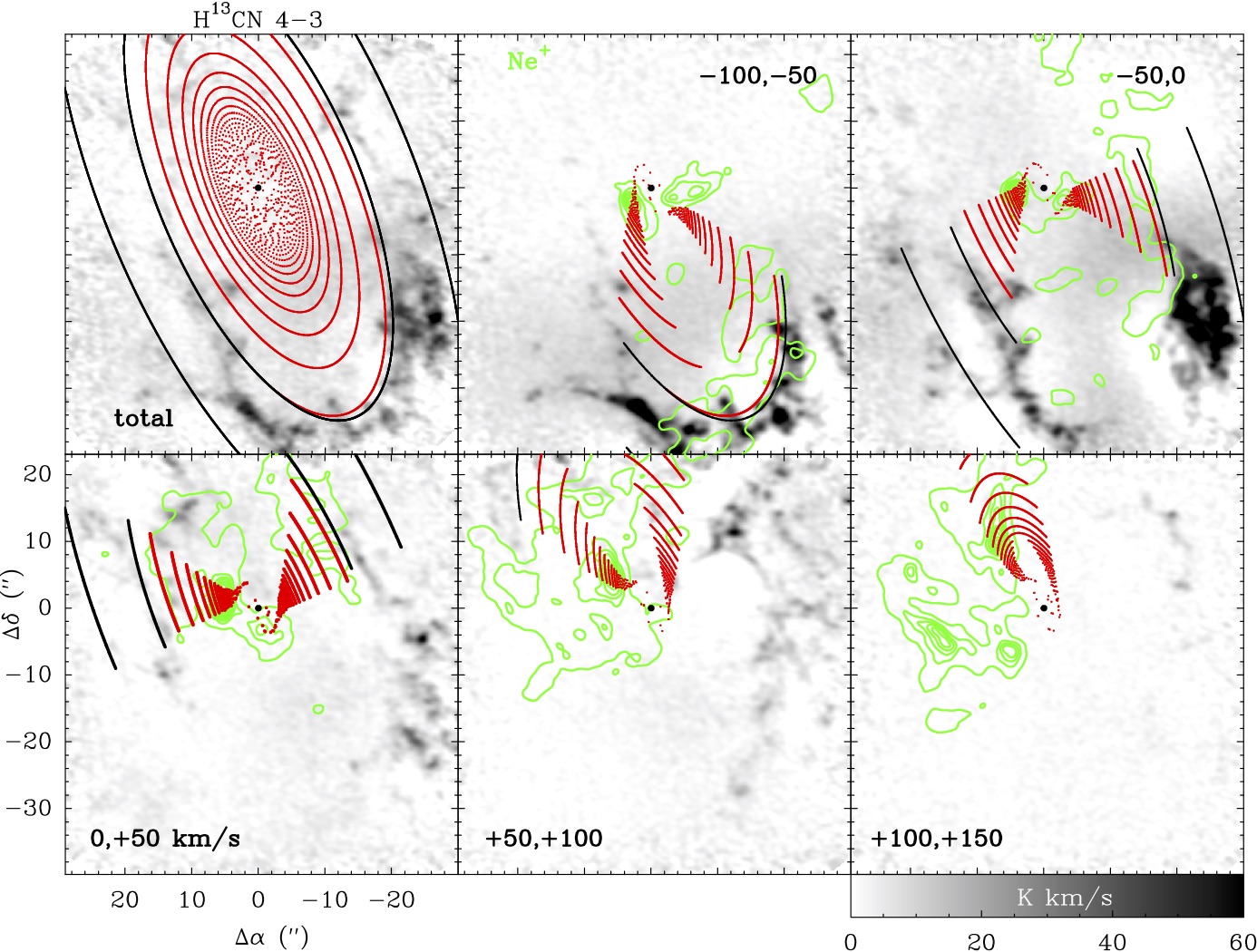}
\hspace{0.3cm}
\includegraphics[scale=0.18,angle=0]{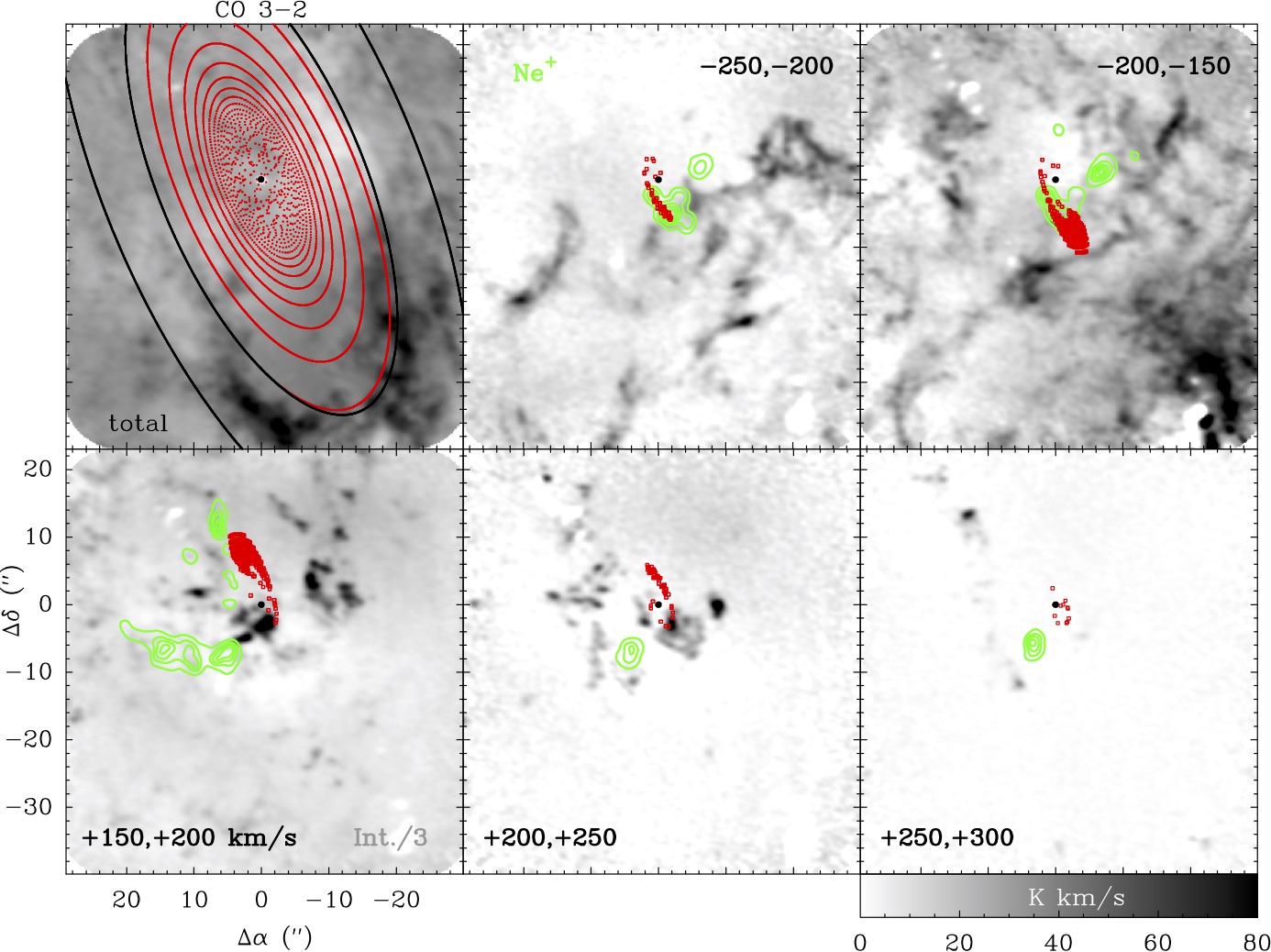}\\
\caption{Modelled spiral orbits (red segments) over H$^{13}$CN~($J$=4-3)  channel maps
at typical CND velocities \mbox{v$_{\rm LSR}$ from $-$100 to $+$150~km\,s$^{-1}$} 
(\textit{left}) and over higher-velocity $^{12}$CO~($J$=3-2) channel maps (\textit{right}), both in steps of 50~km\,s$^{-1}$.
The spirals are  nearly circular orbits ($\iota$=66$^{\circ}$ and $\theta$=23$^{\circ}$) with a small inward radial velocity component  ($a$=$-$0.06\,pc, see text) proposed by
\citet{Irons12} to explain the motion of the ionized gas (shown in green contours) across the the northern arm and western arc.
  Black curves represent a CND model bounded by two rotating circular rings  with inner and outer deprojected radii of 1.5 and 2.3 pc respectively. Sgr\,A$^*$~is  displayed with a black dot.}   
\label{fig:spirals_h13cn_co}
\end{figure*}
%--------------------------------------------------------------------------------

\subsection{Plausible cloudlet motions inside the central cavity  }\label{subsec-discussion-analysis}

We investigate the possible motions of the molecular cloudlets seen toward the central cavity. We start by confronting the spiral and the ellipse  models of the orbiting ionized streamers. We compare  models predictions with the  $^{12}$CO  high-speed  \mbox{$|$v$_{\rm LSR}| > 150$~km\,s$^{-1}$}  emission,
and with the lower velocity range typical of the CND, v$_{\rm LSR}$ from $-$150 to $+$150~km\,s$^{-1}$, for the H$^{13}$CN emission. In some instances, the ionized and the molecular gas are directly related,  for example the western arc and the CND-S, so that this comparative analysis  is justified.

\subsubsection{The spiral model}
This is an extension of the idealized CND circular motion of previous section but the molecular gas can now have  a small inward radial velocity component v$_{\rm R}(r)$ toward Sgr\,A$^*$ (\mbox{Fig.~\ref{fig:circular_orbit}}). 
Hence, the gas originally rotating in the CND 
slowly spirals inwards the cavity until stretched, disrupted, or photo-dissociated and ionized by the strong UV radiation field from the nuclear star cluster. 
The observed gas velocity of such a spiral will be 
\begin{equation}      
\rm {v}_{\rm LSR}(r)=v_{\phi}(r)\,sin\,\iota\,cos\,\phi + v_{\rm R}(r)\,sin\,\iota\,sin\,\phi.
\end{equation}
\citet{Irons12} argue that the kinematics of the ionized gas  in the western arc and northern arm can be explained, as a single structure, if the gas moves along circular orbits with a small radial velocity  \mbox{v$_{\rm R}(r)=(a/r)$\,v$_{\phi}(r)$}, that is, increasing closer
to Sgr\,A$^*$, and resulting in a single one-armed spiral density wave.  
Their best model is for \mbox{$\theta$=23$^{\circ}$}, \mbox{$\iota$=66$^{\circ}$}, and \mbox{$a$=$-$0.06\,pc}, thus implying \mbox{v$_{\rm R}$$\simeq$4~km\,s$^{-1}$} at the inner edge of the CND. 
    In Figure~\ref{fig:spirals_h13cn_co} we show this spiral model applied to different velocity channels of the molecular emission. We just show one spiral, moving in a single plane, as representative of a family of similar orbits.  
Only at low velocities ($|$v$_{\rm LSR}$$|$$<$100~km\,s$^{-1}$; 
left panels in Fig.~\ref{fig:spirals_h13cn_co}) there is 
possible agreement
between some of the observed H$^{13}$CN emission features and the synthetic spiral tracks. These features are consistent with molecular gas leaving the CND and finding its way inward the cavity through nearly circular orbits partially protected from the  strong stellar UV radiation. We propose that the trail of such spirals could contribute to the  more diffuse and extended emission as well.

The timescale required for a  cloudlet to spiral from the inner edge of the CND to 5$''$ from Sgr\,A$^*$ (a~deprojected radius of~0.2~pc) would be $\simeq$1.5$\cdot$10$^5$~yr. Therefore, such a cloudlet will turn several times around Sgr\,A$^*$ before reaching this radius. For molecular cloudlets, such  timescales mean that they will be likely photoevaporated before reaching the very central region unless their column densities
are large enough to provide some UV-shielding (\mbox{see Section~\ref{subsec-analysis-FUV}}). 
Indeed, the spiral model cannot  fit the kinematics of the $^{12}$CO cloudlets  at  extreme velocities \mbox{($|$v$_{\rm LSR}| > 200$~km\,s$^{-1}$)}  because it only predicts such elevated velocities close to Sgr\,A$^*$ and not throughout the  cavity as observed (see right panels in Figure~\ref{fig:spirals_h13cn_co}).
%Instead, many cloudlets are (apparently) distributed along the three  elliptical orbits %\citep[][]{Zhao09} most often invoked to explain the kinematics of the ionized gas %streamers  (Fig.~\ref{fig:co-vchannels}).

\subsubsection{The ellipse model} 
 
\begin{figure}[h]
\centering
\includegraphics[scale=0.7,angle=0]{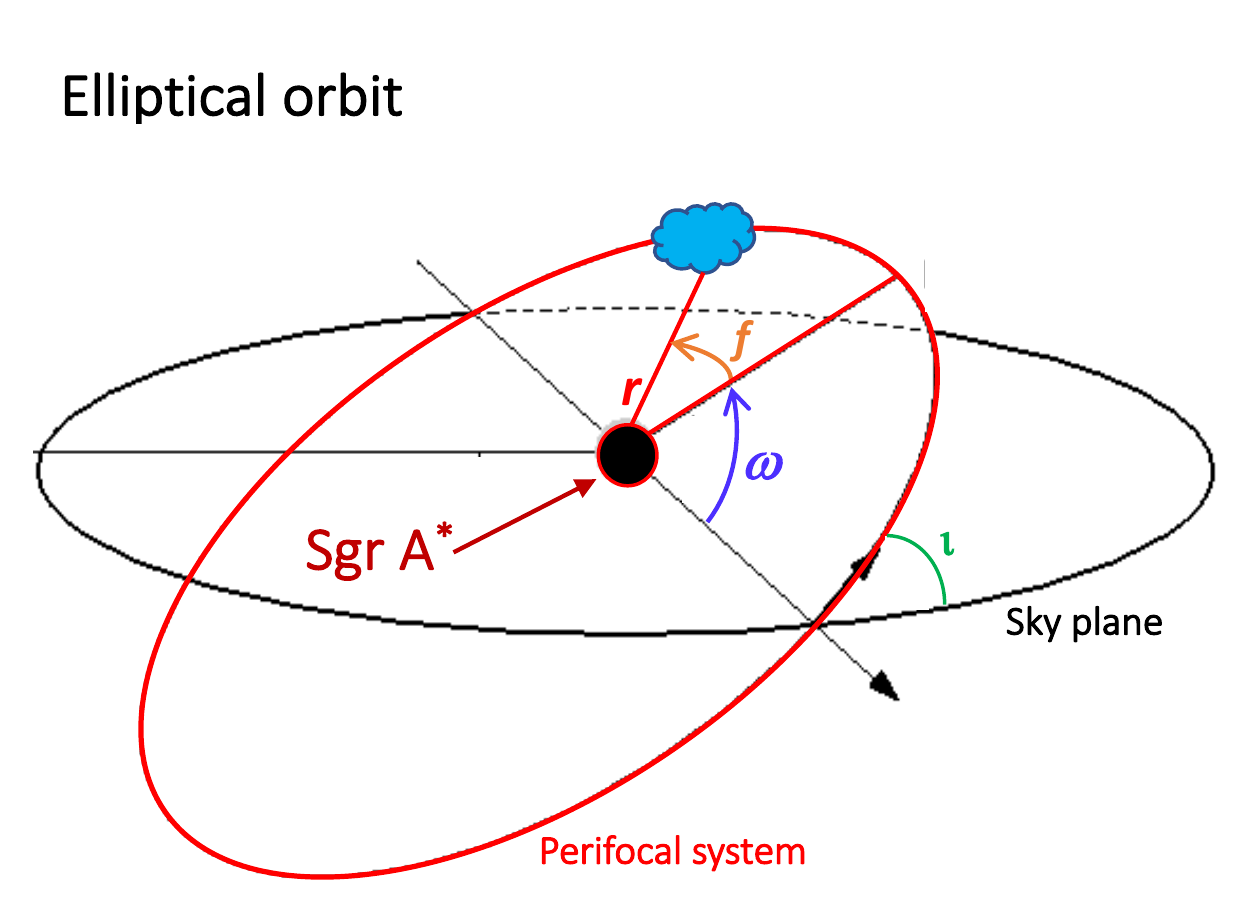}\\
\caption{Cloudlet in an elliptical  orbit.  In the perifocal frame, with Sgr~A$^*$ in one of the ellipse focus, the position the cloudlet is determined by the true anomaly angle $f$. In the plane of the sky, the velocity measured along the line-of-sight depends on the argument of periapsis $\omega$ and
on the inclination angle $\iota$ between the orbit and the sky planes}
\label{fig:ellipse_orbit}
\end{figure}

The line-of-sight velocity of a cloudlet following a Keplerian elliptical orbit with the SMBH at one focus is:   
\begin{equation}\label{eq-ellipses} 
{\rm v}_{\rm LSR}=\sqrt{GM/p}\,[(e + cos\,f)\,cos\,\omega\,sin\,(\pi - \iota)
                 - sin\,f\,sin\,\omega\,sin\,(\pi - \iota)],
\end{equation}   
where $p=a(1-e^2)$  ($a$ is the ellipse semimajor axis and $e$ the eccentricity), $\omega$ is the argument of periapsis, $\iota$ is the orbit inclination angle (defined as in the previous section), and $f$ is the angle of true anomaly defined as 
\mbox{$cos\,f = p/(e \cdot r - 1/e)$} (with $r$ the radial distance from the 
orbiting cloudlet to Sgr\,A$^*$, see  Figure~\ref{fig:ellipse_orbit}). 

%--------------------------------------------------------------------------------
%--------------------------------------------------------------------------------

\begin{figure*}[t]
\centering
\includegraphics[scale=0.18,angle=0]{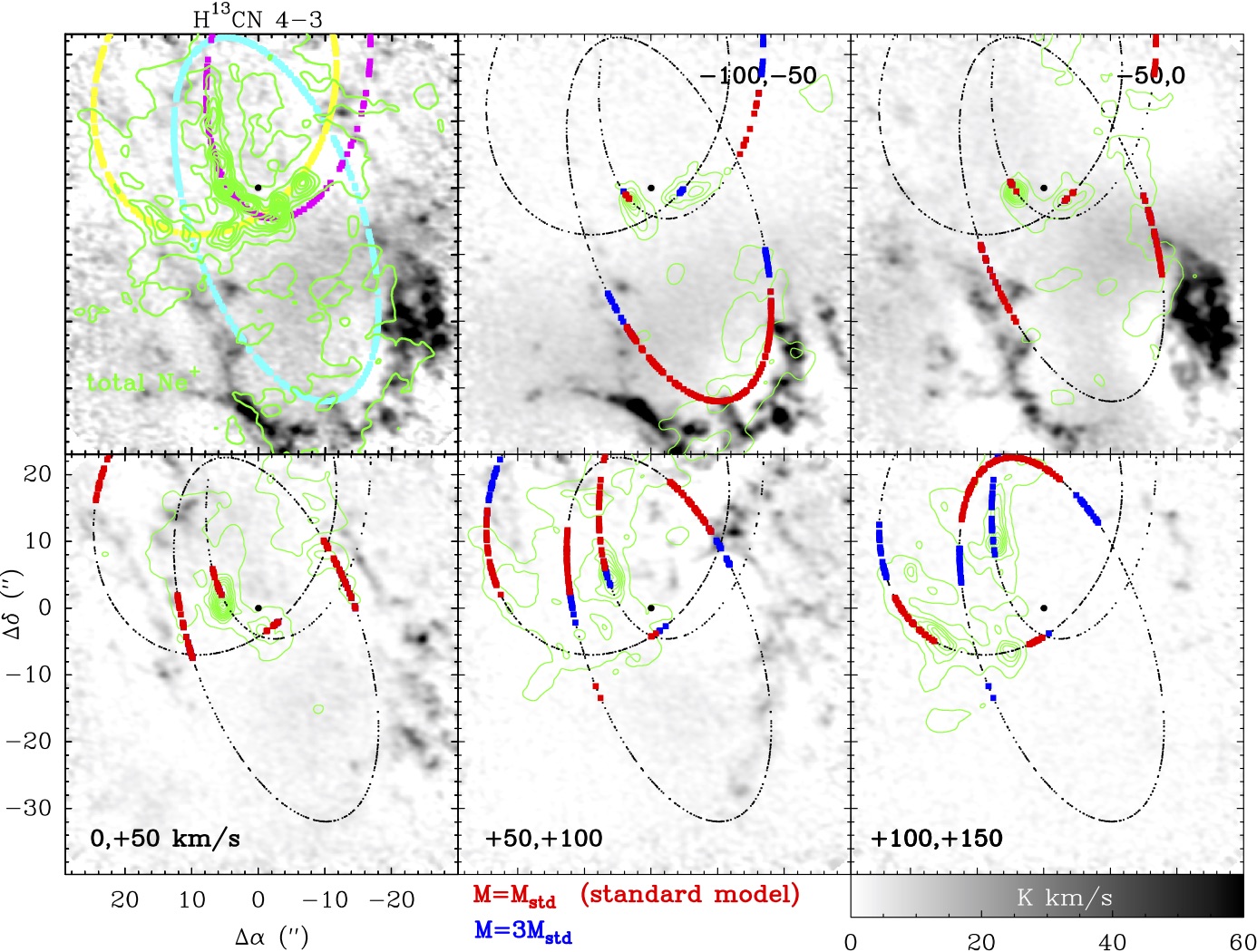}
\hspace{0.3cm}
\includegraphics[scale=0.18,angle=0]{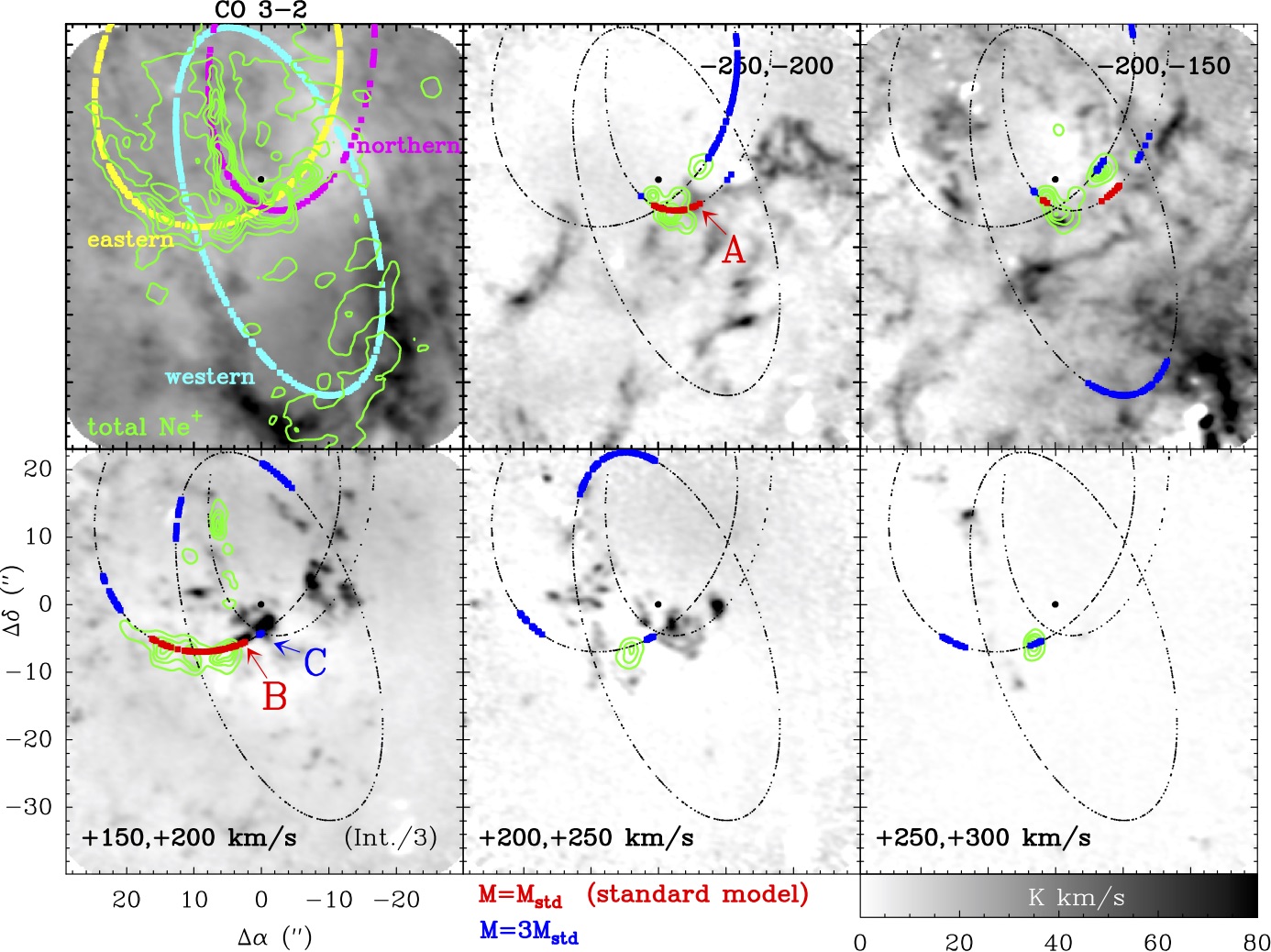}\\
\caption{Modelled Keplerian ellipse orbits over H$^{13}$CN~($J$=4-3) channel maps at typical CND velocities v$_{\rm LSR}$ from $-$100 to $+$150~km\,s$^{-1}$ (\textit{left}) and over higher-velocity $^{12}$CO~($J$=3-2) channel maps (\textit{right}). Green contours show the [\NeII]12.8\,$\mu$m emission from ionized gas streamers \citep{Irons12}. The red orbit segments show the expected  location of the gas according to the best-fit orbital parameters of 
\citet{Zhao09} for the ionized gas (standard mass model). The blue orbit segments are for a model in which the dynamical mass ($M$ in Eq.~2) is multiplied by 3.}   
\label{fig:ellipse_models}
\end{figure*}

The red  segments in Figure~\ref{fig:ellipse_models} show the expected location and v$_{\rm LSR}$ velocities of the
three families of Keplerian  orbits  that better fit the spatial distribution and velocities of the ionized gas streamers, the standard model \citep[see Table~5 of][for the orbital parameters]{Zhao09}. The green contours show the [\NeII]12.8\,$\mu$m emission in the same velocity bins. 
At low negative-velocities, typical of the \mbox{CND-S} lobe, the molecular gas matches the kinematics of the ionized western arc
(chanel maps v$_{\rm LSR}$=$-$100 to 0~km\,s$^{-1}$   in Fig.~\ref{fig:ellipse_models}\,\textit{left}). 
Other than that, there is no other major correspondence between the low velocity H$^{13}$CN 
and the ionized gas emission toward the cavity.

%--------------------------------------------------------------------------------
%--------------------------------------------------------------------------------

The right panels in Figure~\ref{fig:ellipse_models}  show the predictions of the ellipse model, red segments, over the higher velocity \mbox{($|$v$_{\rm LSR}|$$>$150~km\,s$^{-1}$)} $^{12}$CO emission.
Despite the apparent match between the position of several cloudlets along the ellipse orbits, the velocities of most $^{12}$CO cloudlets are not consistent with the velocities of the orbiting ionized streamers. In particular, the HPVG cloudlets seen as a prolongation of the eastern arm, in the north-west direction, appear at high positive-velocities
\mbox{$$v$_{\rm LSR}$=$+$150 to $+$200~km\,s$^{-1}$} channel map, whereas the  model predicts, and the ionized gas is observed at,  negative velocities. Similarly, the EPVG cloudlets seen along the western orbit, \mbox{$$v$_{\rm LSR}$=$+$200 to $+$300~km\,s$^{-1}$} channel maps, appear at much higher positive velocities than those predicted by the model.
Only the molecular structures labelled as A and B, including HPVG cloudlet $\#$1, show agreement in both position and velocity along the northern and eastern Keplerian orbits. In the ellipse model, these orbit positions are located at a deprojected distance of $\simeq$0.3~pc ($\simeq$1~lyr)  from Sgr~A$^*$.

To test the impact of a higher central  dynamical mass in the orbit velocities, in 
\mbox{Figure~\ref{fig:ellipse_models}} we also show a more extreme model in which we increase the enclosed  mass ($M$ in Eq.~2) by a factor of~3, thus increasing the gas velocities
along the orbits (blue segments in  Fig.~\ref{fig:ellipse_models}). Only for the molecular structure C, including HPVG cloudlet $\#$2, the CO emission
 matches the predictions  well. However, according to our current knowledge of the SMBH and stellar cluster masses \citep{Genzel10,Schodel14} this model is less realistic than the standard one.

\subsection{Molecular gas densities, masses and extinction}\label{subsec-analysis-masses}

We estimate the $^{12}$CO-- and H$^{13}$CN--emitting gas column densities and masses 
 over the different velocity intervals. We assume a single rotational temperature ($T_{\rm rot}$) and a \mbox{Boltzmann} distribution of rotational levels. If the abundance of the species $s$ with respect to H nuclei is known (\mbox{$x(s)=N(s)/N_{\rm H}$}) one can estimate the total \mbox{$N_{\rm H}=N$(H+2H$_2$)} column density of molecular gas for each velocity range, and also the visual extinction along a given line-of-sight \mbox{$A_V\simeq N_H\,({\rm cm^{-2}})/1.9\cdot10^{21}$~mag} for standard interstellar grains. We note the much higher critical density (\mbox{$n_{\rm cr}$}) for collisional excitation of H$^{13}$CN rotational transitions  
 compared to those
of CO: \mbox{$n_{\rm cr}$(H$^{13}$CN 4-3)$\simeq$10$^8$~cm$^{-3}$} versus \mbox{$n_{\rm cr}$(CO 3-2)$\simeq$4$\cdot$10$^4$~cm$^{-3}$}. For gas densities 
$n_{\rm H}$$\ll$$n_{\rm cr}$, this leads to subthermal emission, thus low rotational temperatures ($T_{\rm rot}$(H$^{13}$CN)$\ll$$T_{\rm k}$) but it does not necessarily mean the the H$^{13}$CN--emitting gas is very dense, as it is often assumed in the literature of the Galactic center. 

The  H$^{12}$CN abundance  is well constrained by chemical models  to \mbox{$x$(H$^{12}$CN)$\approx$5$\cdot$10$^{-9}$} \citep{Boger05}. Here we take $x$(H$^{13}$CN)=10$^{-8}$/25  to accommodate the higher metallicity and low isotopic ratio 
$^{12}$C/$^{13}$C=25 in the Galactic center \citep{Mezger96,Martin12}. We integrate the column density equations for optically thin line emission
\begin{equation}\label{eq-N}
N_{\rm thin}(s)=8\pi\,\frac{\nu^3}{c^3} \frac{Q(T_{\rm rot})}{g_u \, A_{ul}} 
\frac{1}{1 - e^{E_u/kT_{\rm rot}}}
\frac{\int{T_B({\rm v})\,d{\rm v}}}{J(T_{\rm rot})-J(T_{\rm bg})} \,\,\, {\rm (cm^{-2})},
\end{equation}over different areas and velocity ranges. In this equation \mbox{$J(T)$=$T^*$/($e^{T^*/\,T} - 1$)} is the equivalent brightness temperature, with $T^*$=$E_u/k\,T$. The integral term represents the line intensity in K\,km\,s$^{-1}$ and  $T_{\rm bg}$ is the cosmic microwave background temperature.
If the emission is slightly optically thick, line opacity $\tau_{\rm emi}$$\simeq$1,  one can add an optical depth correction  \mbox{$N(s)$= $N_{\rm emi}(s)\cdot\tau_{\rm emi}/(1-e^{-\tau_{\rm emi}})$}, where                   
\begin{equation}\label{ec-tau_emi}
\tau_{\rm emi}=-{\rm ln}\,\left[1-\frac{T_{\rm B}^{\rm Peak}}{J(T_{\rm rot})-J(T_{\rm bg})}\right], 
\end{equation}
can be estimated from observations. 

%Only toward the CND, the $^{12}$CO emission is $\tau_{\rm emi}$$\gg$1. 

 For the expected range of rotation temperatures\footnote{
 $T_{\rm rot}$(H$^{12}$CN)=43$\pm$6~K from H$^{12}$CN~$J$=6-5 to 8-7 line observations \citep{Goicoechea18}.}  
 $T_{\rm rot}$(H$^{13}$CN)=20 to 100~K, we determine an H$_2$ mass of  \mbox{700-1100~$M_{\odot}$} in the CND-S lobe \mbox{($|$v$_{\rm LSR}|<100$~km\,s$^{-1}$)}.
 For a mean core diameter in the CND of \mbox{0.15-0.25~pc} \citep[see also][]{Christopher05,Lau13}, we estimate an average core gas density of  \mbox{$<n_{\rm H}>$$\simeq$10$^5$~cm$^{-3}$} and density peaks of 
 \mbox{$n_{\rm H}$$\simeq$2$n$(H$_2$)$\simeq$(3-4)$\cdot$10$^5$ cm$^{-3}$}. These values would be a factor of about five higher if  $x$(HCN)$\simeq$2$\cdot$10$^{-9}$, the lower limit of chemical model predictions \citep{Boger05}.

Our inferred masses and densities for the molecular cores in the CND 
roughly agree, actually they are up to a factor ten lower, with  
those derived from local LVG models of  optically thick H$^{12}$CN lines \citep{Mills13,Smith14}.
Our simple estimates from \mbox{H$^{13}$CN ($J$=4-3)} agree more with the densities inferred from the less abundant species HCO$^+$ \citep{Mills13}. This suggests that
for abundant high-dipole moment molecules, in addition to \mbox{mid-IR} pumping, line optical depth effects determine the emergent line intensities and their spatial distribution.
In particular, the  H$^{12}$CN  emission that arises from inside the cores can
be scattered by lower density envelopes and dispersed over larger areas than the true spatial extend of the cores, for example traced by H$^{13}$CN. This is seen
in observations \mbox{\citep[][]{Cernicharo87}} and is explained by more accurate non-local radiative transfer models \citep[][]{Gonzalez93,Goicoechea06}.

In any case, our analysis and the more detailed excitation analyses demonstrate that the densities of the CND cores are lower,
by approximately two orders of magnitude, than the local Roche density: several 10$^7$~cm$^{-3}$.  This supports that the  CND cores are unstable against tidal shear from the central mass
\citep{Gusten87,Jackson93,Requena12,Lau13,Mills13,Smith14}. Hence, they will  disaggregate, creating streamers and ring segments, and have difficulty forming new stars.
This conclusion, however, does not immediately imply that the entire CND is a transient entity itself (see Discussion).  
 
The  mass for the \mbox{CND-S} lobe estimated from H$^{13}$CN, \mbox{$\approx$10$^3$-10$^4$~$M_{\odot}$}  is lower than previous studies that presume that cores are massive enough to be self-gravitating.  In other words, that their broad line-widths, 
\mbox{$\Delta$v$\approx$20-30~km\,s$^{-1}$}, are caused by virialized motions and not by turbulence and  strong tidal shear \mbox{\citep[e.g.,][]{Christopher05,Montero09}}. Instead, they agree more with 
detailed HCN and HCO$^+$ excitation and line radiative transfer analyses \citep{Mills13,Smith14}, and with the dust continuum emission \citep{Telesco96,Latvakoski99,Etxaluze11,Lau13} although grain models also have caveats: composition, size distribution, etc.
 
Following the single $T_{\rm rot}$ method summarized in Eq.~3  we also estimate the column densities associated with the high velocity $^{12}$CO cloudlets detected inside the cavity.  We assume a uniform CO abundance of $x$($^{12}$CO)=2$\cdot$10$^{-5}$ and $T_{\rm rot}$=100~K. This $T_{\rm rot}$ value has been  estimated from spatially \mbox{unresolved} observations high positive-velocity CO~$J$=5 to 10 line emission wings  \citep{Goicoechea18}.
 Under strong far-UV 
(FUV; stellar photons with 
\mbox{$5 < E < 13.6$~eV}) irradiation conditions (\mbox{$G_0\geq 10^4$} in Habing units) state-of-the-art photochemical models
 predict $^{12}$CO abundances ranging from $x$($^{12}$CO)$\simeq$10$^{-6}$ at low visual extinctions \mbox{$A_V$ ($<$1-2~mag)}, to \mbox{$x$($^{12}$CO)$\simeq$10$^{-4}$} in well shielded molecular cloud interiors \citep[e.g.,][]{Joblin18,Goicoechea18}. Here we take the mean value and multiply by 2 to account for the higher metallicity of the Galactic center \citep{Mezger96,Bolatto13}. 
Because we estimate \mbox{$\tau_{\rm emi}$($^{12}$CO $J$=3-2)$\gtrsim$1} toward several cloudlets, we allow for an opacity correction factor 
estimated from the observed line peak temperature (Eq.~\ref{ec-tau_emi}).
Only toward the CND, at lower v$_{\rm LSR}$, the $^{12}$CO emission is $\tau_{\rm emi}$$\gg$1. 

\begin{figure}[t]
\centering
\includegraphics[scale=0.21,angle=0]{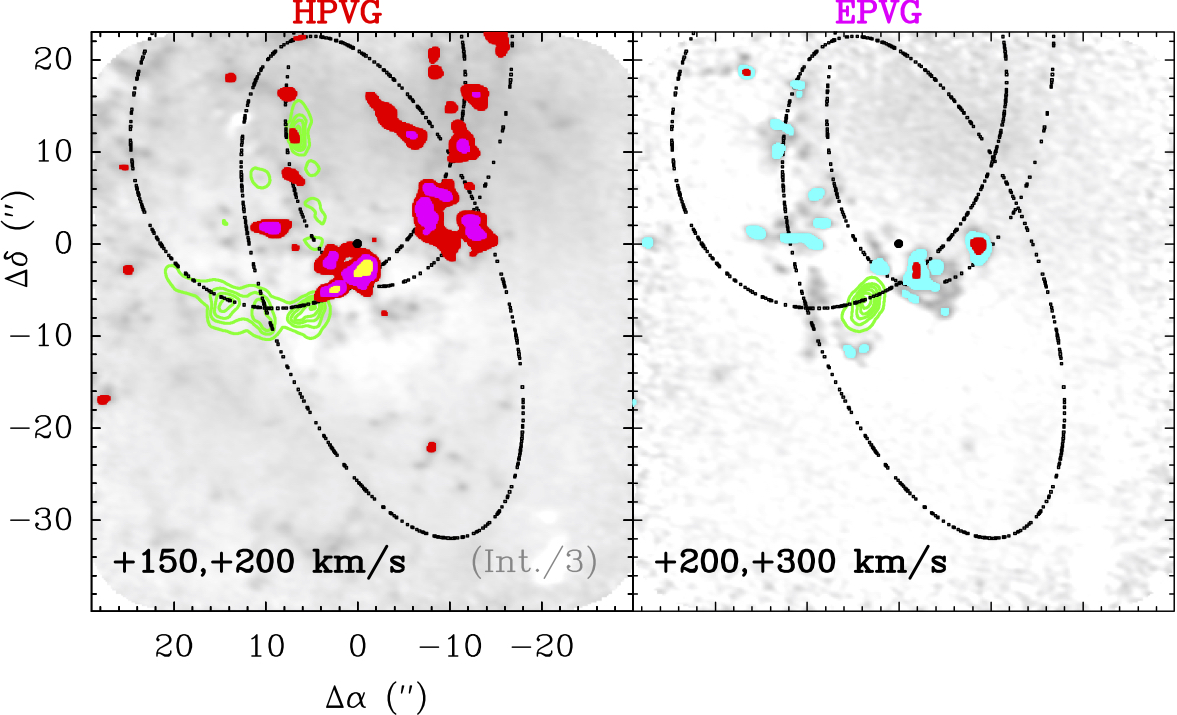}
\caption{Approximate visual extinction ($A_V$) map produced by the HPVG and EPVG cloudlets and obtained
from $^{12}$CO
assuming a uniform abundance $x$($^{12}$CO)=2$\cdot$10$^{-5}$ and a single rotational temperature
$T_{\rm rot}$($^{12}$CO)=100~K \mbox{(see text)}. 
 The color code shows the estimated extinction peaks: \mbox{$A_V$$\approx$6-9 (yellow)}, \mbox{$A_V$$\approx$3-6 (magenta)}, \mbox{$A_V$$\approx$1.5-3 (red)}, and \mbox{$A_V$$\approx$0.6-1.5~mag} (cyan, shown only in the right panel). 
The gray images show the $^{12}$CO~$J$=3-2  line emission integrated in the velocity interval displayed in each panel. The green contours shows the  ionized gas [\NeII]12.8\,$\mu$m emission at the same velocities \citep[][]{Irons12}.  }
\label{fig:extinction}
\end{figure}

For the EPVG component (v$_{\rm LSR}$=$+$200 to $+$300~km\,s$^{-1}$) we derive extinction peaks of $A_V$$\simeq$1.5-1.8~mag toward 3 EPVG cloudlets 
 and a total H$_2$ mass of 6-12~$M_{\odot}$. This value includes the $^{12}$CO EPVG cloudlets and the more extended diffuse emission component, $A_V$$\simeq$0.1-0.3~mag, detected at these velocities.
      The typical cloudlet sizes are roughly 2$''$. This translates into a few gas density peaks of $n_{\rm H}$$\simeq$(1.2-2.5)$\cdot$10$^4$~cm$^{-3}$, thus very far from the critical density, 
\mbox{$n_{\rm Roche}\simeq6\cdot10^{10}\,(0.1\,{\rm pc}/r)^3$} (cm$^{-3}$) needed to overcome the tidal shear in the  central pc \citep{Morris96,Genzel10}. These cloudlets are thus stretched  and many of them violently disrupted into smaller constituents of lower mass and $A_V$.
Figure~\ref{fig:extinction} (right) shows a map of the approximate visual extinction
produced by the EPVG cloudlets obtained by assuming a uniform $^{12}$CO abundance and $T_{\rm rot}$.

 %--------------------------------------------------------------------------------
%--------------------------------------------------------------------------------

A similar analysis of the velocity range v$_{\rm LSR}$=$+$150 to $+$200~km\,s$^{-1}$ leads to an average visual extinction value of \mbox{0.6-1.0~mag} through the field, and several   $^{12}$CO HPVG cloudlets with visual extinction peaks of \mbox{$A_V$$\simeq$3-9~mag} (Fig.~\ref{fig:extinction} left). This implies the presence of a few cloudlets with enough shielding  against strong FUV radiation fields. The estimated H$_2$ mass in this velocity interval is 45-90~$M_{\odot}$, but it includes the possible contribution from background emission sources not related to Sgr\,A$^*$
(see Section~\ref{subsec-analysis-absorption}).
%%: the ``expanding molecular ring'' or EMR \citep[e.g.,][, see Section~\ref{subsec-analysis-absorption}]{Scoville72}. 
 Our estimated extinction and mass values inside the cavity have to be multiplied by about two  if $T_{\rm rot}(\rm CO)=$200~K and reduced by a factor five if the assumed CO abundance is increased to $\sim$10$^{-4}$.

\subsection{FUV attenuation and photo-dissociation timescales}\label{subsec-analysis-FUV}

The near-infrared (NIR) extinction toward Sgr\,A$^*$ is $A_{Ks}$$\approx$3 \citep[$\lambda_{Ks}$=2.2~$\mu$m,][]{Genzel10,Schodel10}.  
This value includes \mbox{$A_V$$\approx$30 magnitudes} of visual extinction ($\lambda_{V}$=0.55~$\mu$m) produced by cold foreground clouds
in the line-of-sight toward the Galactic center \citep[e.g.,][]{Moneti01}.
Combining Paschen-$\alpha$ line ($\lambda_{P\alpha}$=1.87~$\mu$m), 6~cm radiocontinuum, and \mbox{mid-infrared} (MIR) dust emission observations, \citet{Lau13} presented maps of the  MIR extinction ($\lambda_{MIR}$=19.7~$\mu$m) intrinsic to the CND and cavity. In these maps it is evident that the local MIR extinction is greater along the western arc than toward the central cavity: it peaks at the CND-S lobe,
it is low along the ionized streamers, and it rises again close to Sgr~A$^*$ (north-west).

%The absolute extinction can be reconciled if the dust and molecular gas in the CND is %clumpy \citep{Lau13}. This agrees with the spatial distribution of the H$^{13}$CN %($J$=4-3) emission
%(Fig.~{fig:h13cn})-   

\citet{Ciurlo16} presented a $\sim$1$''$-resolution  NIR extinction map toward the innermost $\simeq$30$''$$\times$30$''$ region obtained from spectrally-unresolved H$_2$ observations. They derive an averaged value of $A_{Ks}$$\simeq$3.4 and maxima
around $A_{Ks}$$\simeq$4,  thus allowing for local peaks of several
magnitudes of visual extinction from dust and gas inside the cavity itself. 
In particular, some of the infrared extinction peaks outside the mini-spiral 
coincide with the position of some $^{12}$CO HPVG cloudlets.
Therefore, at least qualitatively, the presence of small-scale molecular cloudlets of a few magnitudes of visual extinction inside the cavity, partially shielded from the strong FUV field, does not contradict the existing
extinction maps.

O-type massive stars in the central cluster \citep[][and references therein]{Becklin68,Krabbe95,Schodel14} produce strong UV radiation fields and launch powerful winds \citep[e.g.,][]{Wolfire90,Mezger96,Genzel10}. FUV photons heat the 
interstellar dust to high temperatures,  above $\sim$100~K. 
Hence, MIR and FIR continuum observations  of the dust thermal emission \citep{Telesco96,Latvakoski99,Etxaluze11,Lau13} can be used to estimate the FUV-photon flux ($G_0$). The 19.7/31.7~$\mu$m color temperature ($T_{\rm col}$) peaks at $T_{\rm d}$$\approx$145~K inside the cavity, and is $\approx$85~K at the inner  edge of the CND-S lobe \citep[][]{Lau13}. 
We use $G_0$$\simeq$($T_{\rm d}$/12.2)$^5$  \citep[see][]{Hollenbach91}, where we adopt \mbox{$T_{\rm d}$\,(big grains)\,$\simeq$\,$T_{\rm col}$}  and assume that the MIR continuum  is dominated by  the thermal emission of \mbox{FUV-heated} big grains, these color temperatures imply a FUV flux ranging from $G_{\rm 0,Cavity}$$\approx$10$^{5.3}$ to 
$G_{\rm 0,CND}$$\approx$10$^{4.2}$. %%Given these high FUV-radiation doses, 
The assumption  that the observed MIR  continuum emission is dominated by  big grains, and not by stochastically heated very small grains (VSGs) or PAHs  out of thermal equilibrium \citep[i.e.,][]{Desert90},
is a reasonable one because the thermal emission peak from hot  dust 
\mbox($T_{\rm d}$=150-75~K)  lies  at MIR wavelengths. 
Hence, this emission will dominate over that of other grain populations. Still, our estimated
$G_{\rm 0}$ values should be considered as upper limits because
of the undetermined contribution of VSGs and PAHs to
 the observed MIR photometric emission
\citep[e.g.,][]{Berne07,Etxaluze11,Etxaluze13}. 

\begin{figure}[t]
\centering
\includegraphics[scale=0.15,angle=0]{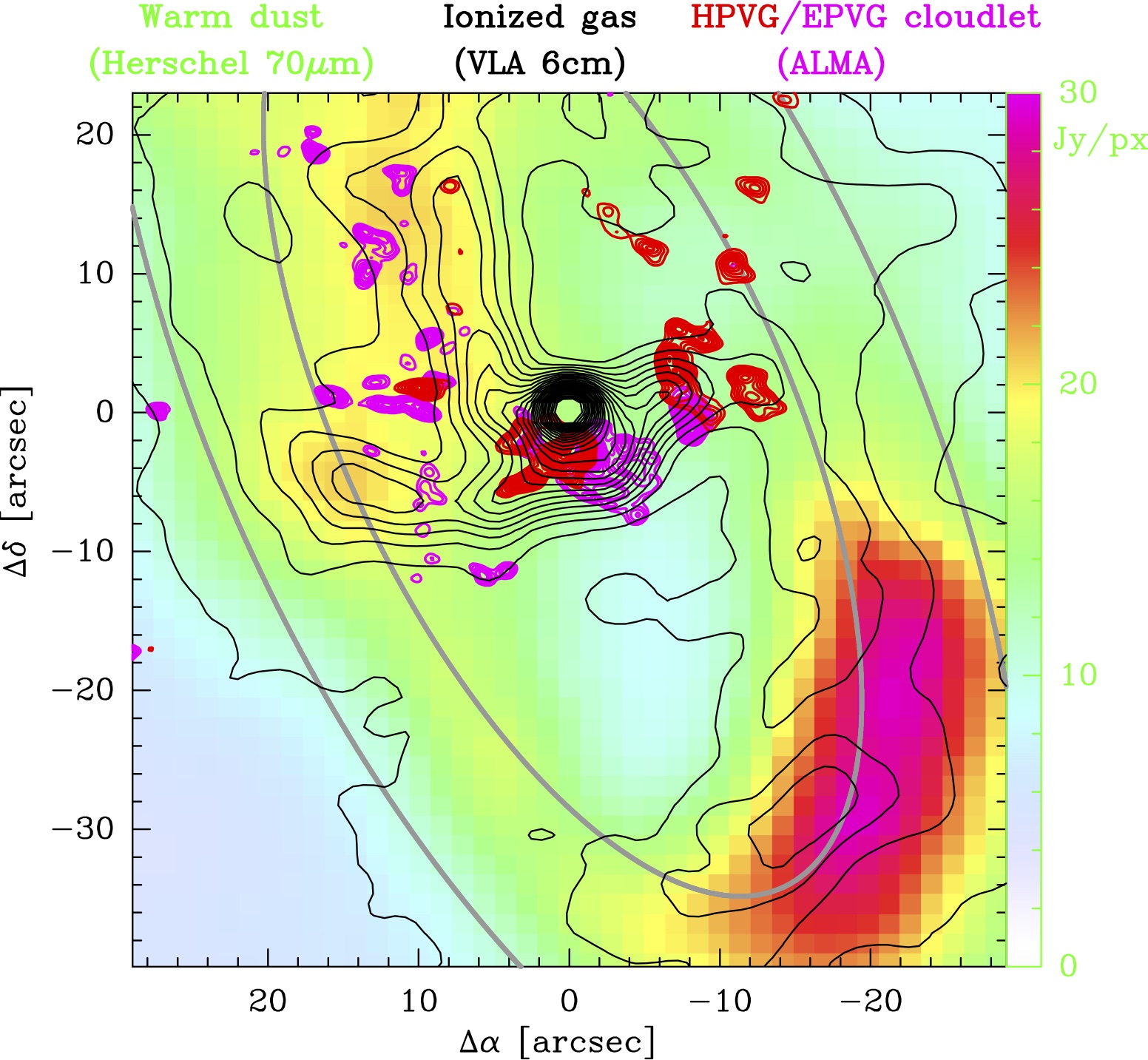}
\caption{Location of the HPVG (red) and EPVG (magenta) $^{12}$CO cloudlets detected by ALMA over a color image of the warm \mbox{(FUV-irradiated)} dust emission taken by
Herschel/PACS at 70~$\mu$m. Black contours show the ionized gas streamers observed by VLA in  6~cm radio-continuum emission \mbox{\citep{Yusef87}}.}
\label{fig:pacs70-alma}
\end{figure}

Figure~\ref{fig:pacs70-alma} shows the location of the HPVG and EPVG molecular cloudlets with respect to the 70~$\mu$m dust thermal emission 
\citep[observed by \textit{Herschel},][]{Molinari11} typically associated with the presence of FUV-irradiated hot/warm neutral material. 
The 70~$\mu$m dust emission is slightly shifted to the east with respect to the ionized gas in the northern arm, as if this streamer (traced by the 6~cm radiocontinuum emission in Fig.~\ref{fig:pacs70-alma}) was the limb-brightened ionized rim of a roughly edge-on neutral gas structure behind. Several EPVG cloudlets appear distributed along this northern area of bright 70~$\mu$m dust emission that also coincides with the so-called northern intruder seen in MIR dust emission \citep{Telesco96,Latvakoski99}, neutral oxygen [\OI]\,63\,$\mu$m  \citep[][]{Jackson93}, Br\,$\gamma$ \citep{Paumard04} and in 
H$_2$ line emission \citep{Ciurlo16}.

Most EPVG cloudlets appear distributed  along areas of high 19.7/31.7~$\mu$m color temperature ratio \citep[][]{Lau13} and bright 70~$\mu$m dust emission inside the cavity. If their true location is not very different from the dust-emitting regions, the molecular cloudlets are certainly irradiated by strong \mbox{FUV-radiation fields} \citep{Wolfire90} and heated, predominantly by shocks, to temperatures of several hundred~K \citep[e.g.,][]{Mills17,Goicoechea13,Goicoechea18}.  
The HPVG cloudlets, are distributed along areas of weaker  70~$\mu$m dust emission, representative of lower FUV-irradiation conditions or lower column
densities of neutral material.

The  attenuation of the FUV-field from the central cavity to the inner edge of the CND-S lobe implies a dust column density
 equivalent to \mbox{$A_V$$\approx$1-2~mag} of local visual extinction, where 
\begin{equation}
A_V\approx -(1/\kappa_{\rm FUV})/(\tau_{\rm FUV}/\tau_V)\,\,{\rm ln}\,
 (G_{0,\rm CND}/G_{0,\rm Cavity}).
\end{equation}
 In the above relation,
$\tau_{\rm FUV}/\tau_V$ is the extinction curve at FUV wavelengths and the $\kappa_{\rm FUV}$ parameter ($\approx$1) depends on grain properties \citep{Goicoechea07} and metallicity \citep{Mezger96}. Hence, the approximate extinction values derived from Eq.~(5)  are consistent with the average values estimated from the 
$^{12}$CO cloudlet emission in the previous section.

%--------------------------------------------------------------------------------
%--------------------------------------------------------------------------------

 Taking a representative FUV photon flux of $G_0$=10$^{4.7}$ and using updated $^{12}$CO photodissociation rates \citep{Visser09}, we derive that the photodissociation timescale ($\tau_{\rm ph}$) of cloudlets within the cavity will be short: 
 $\approx$10$^5$, 2$\cdot$10$^4$, and a few hundred years for cloudlets with  visual extinction peaks $A_V$=5, 4, and 3~mag respectively. 
In addition, we derive the extinction that would allow the molecular gas in the inner edge of the CND to perform one complete turn before being photodissociated. The $A_V$ satisfying $\tau_{\rm rot}(\varphi)=\tau_{\rm ph}$($^{12}$CO) for 
$G_{\rm 0, CND}$=10$^{4.2}$ amounts to $A_V$$\approx$4~mag of visual extinction.
 This value argues in favor of rather low column densities of material at the
inner  CND-S edge, and thus  consistent with the relatively low inferred masses.   
       
\subsection{Absorption spectrum: foreground diffuse gas}\label{subsec-analysis-absorption}

Sgr\,A$^*$ is a strong synchrotron submillimeter emitter, with a continuum emission that peaks around 350~GHz \citep{Marrone06,Eckart12,Moser17}.  For a given continuum equivalent brightness temperature $T_{\rm c}$, any foreground molecular gas  component in the line-of-sight will be observed in absorption if $T_{\rm rot} < T_{\rm c}$. 
The central continuum source %%%%(with $T_{\rm c}$$\simeq$40~K at 345~GHz) 
has a
compact size, thus a pencil-beam spectrum toward Sgr~A$^*$ can reveal these components.
 
\begin{figure}[t]
\centering
\includegraphics[scale=0.57,angle=0]{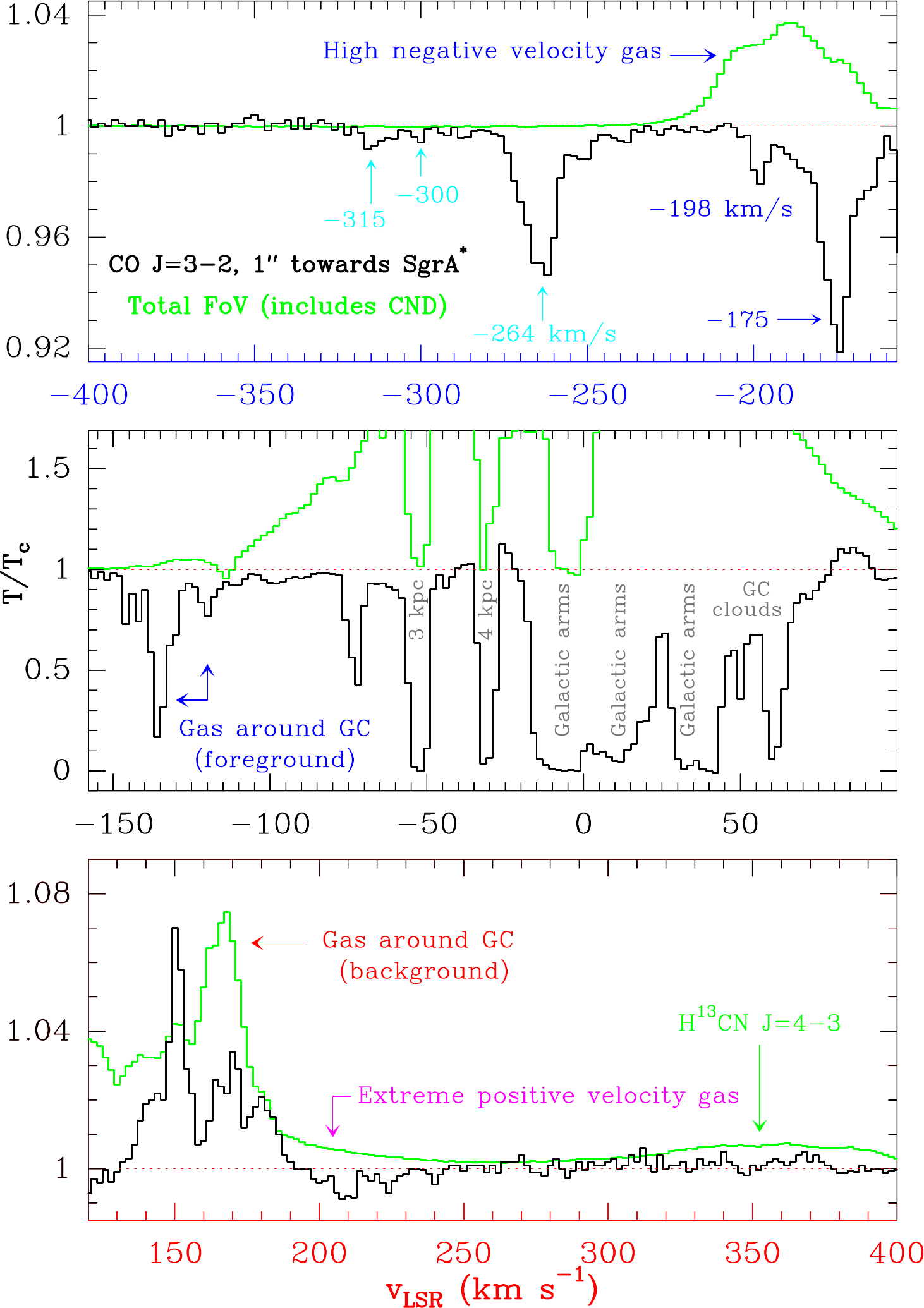}\\
\caption{$^{12}$CO~($J$=3-2) spectrum toward Sgr\,A$^*$. The three panels cover the LSR velocity range from $-$400 to $+$400~km\,s$^{-1}$. The black histogram (dominated by line absorption) is the 1$''$-angular resolution spectrum toward the bright continuum source at Sgr\,A$^*$. The green histogram (dominated by line emission) is the average spectrum over the $\simeq$60$''$$\times$60$''$ field-of-view observed by ALMA.}
\label{fig:absorption-spectrum}
\end{figure}

Because CO is the most abundant molecule after H$_2$,  CO absorption measurements are a very convenient probe of small column densities of low excitation molecular gas, that is, low $T_{\rm rot}$ but not necessarily low $T_{\rm k}$ if gas densities are very low and $T_{\rm c}$ is high.
\mbox{Figure~\ref{fig:absorption-spectrum}} shows the
1$''$-resolution $^{12}$CO~($J$=3-2)  spectrum taken with the ALMA interferometer alone, no zero-spacing data added, toward  Sgr\,A$^*$ 
(black histogram). Many $^{12}$CO spectral features are detected in absorption. This spectrum provides a \mbox{radiography} of the low excitation gas in the line-of-sight toward the Galactic center and of diffuse gas potentially close to Sgr~A$^*$ (see \mbox{Table~\ref{table:absorptions}} 
for the parameters obtained for each  feature).
SO~and H$^{13}$CN are much less abundant  than CO, in other words, much lower columns of absorbing material, and their rotationally excited levels are more difficult to populate in low density gas. This explains the lack of  \mbox{SO~8$_9$-7$_8$}  and
 \mbox{H$^{13}$CN~$J$=4-3}  line absorption toward the continuum emission of Sgr~A$^*$.

Figure~\ref{fig:absorption-spectrum} also shows the \mbox{$^{12}$CO~($J$=3-2)} spectrum 
(green) averaged over the entire ALMA field, and adding the zero-spacing data. This other spectrum is dominated by 
 CO emission from the Galactic center. Comparing both spectra allows one to better understand the velocity components that more certainly arise from gas close to Sgr\,A$^*$ in the ALMA images.

    The velocity range $|$v$_{\rm LSR}|<$100~km\,s$^{-1}$ has been widely studied in absorption measurements\footnote{We  checked that no line from a molecule different than CO  could be responsible of the observed spectral features.
Only the absorption feature at $-121\pm2$~km\,s$^{-1}$ may be blended with the 
\mbox{SH$^+$ 1$_0$-0$_1$ $F$=1/2-3/2} line at 345.944~GHz \citep{Muller14}  if the intrinsic SH$^+$ absorption toward Sgr\,A$^*$ peaks at v$_{\rm LSR}$$\simeq$$+$7~km\,s$^{-1}$. 
The weaker SH$^+$ fine structure component 
\mbox{$F$=1/2-1/2} at 345.858~GHz would then be the small absorption peak at
 $\simeq$$-$46~km\,s$^{-1}$. Although SH$^+$ absorption lines are known to exist toward the Galactic center \citep[e.g.,][]{Godard12} and even if the CO absorption peak at  $-121\pm2$~km\,s$^{-1}$ is blended, we don't expect detectable SH$^+$ emission in the ALMA images; for example in the Orion Bar, the $^{12}$CO~($J$=3-2) line is $>$500 times brighter that the 
 SH$^+$ 1$_0$-0$_1$ $F$=1/2-3/2 line \citep{Goicoechea17}.}. 
 It includes absorptions produced by low-density molecular clouds in the spiral arms of the Galaxy: the  \mbox{3-kpc arm}, \mbox{4-kpc arm}, Sagittarius, and Scutum arms, and by clouds in the Galactic center region \citep{Greaves94,Sandqvist15}. 
The ALMA spectrum also shows  an absorption feature at  \mbox{v$_{\rm LSR}$$\simeq$$-$73~km\,s$^{-1}$} that is not present in \mbox{single-dish},
low angular resolution, absorption spectra \citep[see e.g. in  H$_2$O ground-state line measurements,][]{Sandqvist03}. The feature has a considerable absorption depth, 
\mbox{$\tau_{\rm abs,\,CO\,3-2}$$\simeq$$-${\rm ln}\,($T/T_{\rm c}$)$\simeq$0.8}, equivalent to  $N$(CO)$\approx$2$\cdot$10$^{17}$~cm$^{-2}$ (or $N$(H$_2$)$\approx$5$\cdot$10$^{21}$~cm$^{-2}$) 
for the absorbing material (see below).
VLA observations show an  OH absorption feature at \mbox{v$_{\rm LSR}$$\simeq$$-$73~km\,s$^{-1}$} as well, but only toward the \mbox{CND-S} lobe \citep{Karlsson15}.
The fact that we detect it toward Sgr~A$^*$ implies that it is more extended
around the CND.
 Table~\ref{table:absorptions} shows our estimated column densities of absorbing CO
calculated by integrating each velocity component:
\begin{equation}
N_{\rm abs}(\rm CO) \approx 4.81 \cdot 10^{16} \int{\tau_{\rm abs,\,CO\,3-2}({\rm v})\,d{\rm v}},
\end{equation}
where we have assumed $T_{\rm rot}$=5~K. To derive the equivalent $N$(H$_2$) columns we use $N$(CO)/$N$(H$_2$)=4$\cdot$10$^{-5}$ (see discussion in 
\mbox{Section~\ref{subsec-analysis-masses}}). These column densities would be a factor $\sim$2 higher if one assumes a more elevated $T_{\rm rot}$=10~K.

%--------------------------------------------------------------------------------
%--------------------------------------------------------------------------------

As expected, the $^{12}$CO  spectrum shows a deep absorption feature at 
v$_{\rm LSR}$$\simeq$$-$135~km\,s$^{-1}$ produced by
foreground gas that follows, driven by the barred potential of the Galaxy, \mbox{non-circular} $x_1$ orbits around the Galactic center and produces the characteristic  parallelogram on ($l,{\rm v}$) \mbox{longitud-velocity}  CO maps 
\citep[e.g.,][]{Bally88,Binney91}. This component was first interpreted, and is still often referred to, as an \mbox{expanding molecular ring} \citep[][]{Scoville72,Liszt78}. 
The observed weaker absorptions at v$_{\rm LSR}$$\simeq$$-$121 and $-$145~km\,s$^{-1}$ are likely part of the same widespread structure, that is, molecular gas located in the inner few hundred parsecs of the Galaxy but not associated with the Sgr~A complex. 
In addition, there are features at more negative velocities. Some of these absorptions, those at v$_{\rm LSR}$$\simeq$$-$175 and $-$198 km\,s$^{-1}$, coincide with the so-called $-$180 km s$^{-1}$ feature \citep{Gusten81,Liszt93}. This absorption component arises from HNVG clouds, most likely a  massive cloud falling toward the Galactic center area \citep{Zhao95}.  
Coincidence or not, the \mbox{$-$180 km s$^{-1}$} velocity channel  map  in \mbox{Figure~\ref{fig:co-vchannels}} shows that the emission counterpart resembles a stretched cloud distorted by a central gravitational force 
\cite[see model by][]{Zhao95}. 
\mbox{\citet{Marshall94}} detected, in low resolution \mbox{single-dish} observations, extended \mbox{$^{12}$CO~$J$=3-2} and 2-1 line emission at these negative velocities. From the measured low \mbox{3-2/2-1} line intensity ratio they estimated cold rotational temperatures, 
\mbox{$T_{\rm rot}$=8-10~K\,$\lesssim T_{\rm k}$}, typical of quiescent cold clouds not heated by FUV photons from nearby massive stars. They 
associate the  HNVG  component to the large-scale galactic  gas distribution and not with the environment around Sgr\,A$^*$.
In the ALMA images, this velocity range shows many filamentary structures through the field (Fig.~\ref{fig:co-vchannels}). Hence, owing to foreground structures at \mbox{v$_{\rm LSR}$$\simeq$$-$200 to $-$150~km\,s$^{-1}$}, one cannot easily associate the cloudlets detected in this negative range with Sgr\,A$^*$ surroundings.

\begin{table*}[]
\caption{\label{table:absorptions}
CO ($J$=3-2) line absorption velocity components toward Sgr\,A$^*$ continuum source and estimated column densities.} 
\centering
\begin{tabular}{cccccl}
\hline\hline
v$_{\rm LSR}$$^a$ (km\,s$^{-1}$) & Width$^a$, $\Delta v$ (km\,s$^{-1}$)  &  Depth, $\tau_{\rm abs,\,CO\,3-2}$  &  $N_{\rm abs}$(CO)$^b$ (cm$^{-2}$) & $N$(H$_2$)$^c$ (cm$^{-2}$)  & Component$^d$ \\\hline
-315$\pm$3 &  3$\pm$1 & 0.009  &  1.27$\cdot$10$^{15}$ & 3.17$\cdot$10$^{19}$  & 	\textbf{New}. ENVG. Near Sgr\,A$^*$? \\
-264$\pm$1 & 11$\pm$1 & 0.053  &  2.89$\cdot$10$^{16}$ & 7.23$\cdot$10$^{20}$  & 	\textbf{New}. ENVG. Near Sgr\,A$^*$? \\
-198$\pm$1 & 5$\pm$3  & 0.021  &  4.83$\cdot$10$^{15}$ & 1.21$\cdot$10$^{20}$  &	HNVG. Foreground cloud  \\
-175$\pm$1 & 11$\pm$1 & 0.086  &  4.76$\cdot$10$^{16}$ & 1.19$\cdot$10$^{21}$  &	HNVG. Foreground cloud \\
-145$\pm$2 &  4$\pm$1 & 0.269  &  4.91$\cdot$10$^{16}$ & 1.23$\cdot$10$^{21}$  &    Foreground gas  around GC?\\
-135$\pm$2 &  7$\pm$2 & 1.309  &  4.33$\cdot$10$^{17}$ & 1.08$\cdot$10$^{22}$  &	
Foreground gas  around GC\\
-121$\pm$2 &  6$\pm$2 & 0.269  &  8.48$\cdot$10$^{16}$ & 2.12$\cdot$10$^{21}$  &  Foreground gas  around GC?\\ %SH$^+$? 
-73$\pm$1  &  5$\pm$1 & 0.842  &  2.09$\cdot$10$^{17}$ & 5.22$\cdot$10$^{21}$  &	 CND? \\
-53$\pm$1  &  6$\pm$1 & $\gg$1 &  --- 				   & ---  & 3-kpc arm feature \\
-31$\pm$1  &  5$\pm$1 & $\gg$1 &  ---	               & ---  &	-30 km/s or 4-kpc arm \\ 
-7$\pm$1   & 19$\pm$1 & $\gg$1 &  ---				   & ---  &	Galactic arms \\
12$\pm$1   & 16$\pm$1 & $\gg$1 &  ---                  & ---  &	Galactic arms \\
37$\pm$1   & 21$\pm$1 & $\gg$1 &  ---                  & ---  &	GC clouds \\
50$\pm$1   &  5$\pm$1 & 0.734  & 1.76$\cdot$10$^{17}$  & 4.41$\cdot$10$^{21}$  &	GC clouds \\
61$\pm$1   &  8$\pm$1 & 2.523  & 1.01$\cdot$10$^{18}$  & 2.54$\cdot$10$^{22}$  &	GC clouds \\\hline
\end{tabular}
\tablefoot{$^a$From a multigaussian fit to the 1$''$-resolution spectrum toward
Sgr\,A$^*$.  $^b$Assuming $T_{\rm rot}$(CO)$=$5~K. $^c$Assuming $N$(CO)/$N$(H$_2$)=4$\cdot$10$^{-5}$.  
$^d$ENVG=extreme negative velocity gas, HNVG=high negative velocity gas, GC=Galactic center,
CND=circumnuclear disk.}
%\textbf{References.}...}
\end{table*}

      Surprisingly, the $^{12}$CO absorption spectrum in Figure~\ref{fig:absorption-spectrum} shows two additional weak features at more extreme negative-velocities (v$_{\rm LSR}$=$-$264 and $-$310~km\,s$^{-1}$) that have not been reported before \citep[e.g.,][]{Zhao95,Sandqvist03,Godard12}. 
We term these absorption features at v$_{\rm LSR}$$<$$-$250\,km~s$^{-1}$  the
\mbox{\textit{extreme negative-velocity gas}} (ENVG) component.
      These velocities do not show emission counterpart in the ALMA images either. One interpretation is that they are produced by low density diffuse molecular gas in front of and close to  Sgr\,A$^*$. 
      Given the enclosed mass model and expected velocities, we estimate that they would be located at about \mbox{0.04-0.1~pc} \mbox{(50-120~light-day)} if they follow trajectories with the same inclination as the CND.

The background molecular gas following $x_1$ orbits around the Galactic center,
   that is, the positive-velocity side of the \mbox{($l,{\rm v}$)} parallelogram,
is expected to be seen in CO emission \mbox{(at v$_{\rm LSR}$$\simeq$$+$150-165~km\,s$^{-1}$) }  because it is located behind the continuum source at Sgr~A$^*$  \mbox{\citep{Binney91}}. The ALMA images (Fig.~\ref{fig:co-vchannels}) show both an extended emission, likely dominated by emission from this background component, and several  clumpy features  that we attribute to gas closer to Sgr\,A$^*$. 

Finally, the $^{12}$CO spectrum
averaged over the field shows an emission wing up to 
v$_{\rm LSR}$$\simeq$$+$300~km\,s$^{-1}$.
This  wing is also detected in deep single-dish observations
of rotationally excited \mbox{mid-$J$} CO lines toward Sgr~A$^*$ \citep{Goicoechea18}. At these extreme velocities, the ALMA images unveil a collection of  $^{12}$CO EPVG cloudlets.

\section{Discussion}
\label{sec-discussion}

\subsection{Most cloudlets inside the cavity do not fit the kinematics of the ionized gas streamers}    
 
 In addition to the small spatial-scale molecular emission resolved toward the
 central cavity, perhaps the most relevant result of this work is that the bulk of high  and extreme positive-velocity  $^{12}$CO cloudlets
 (\mbox{v$_{\rm LSR}$$\approx$$+$150 to $+$300~km\,s$^{-1}$})
  do not fit any of the  kinematical models proposed for the ionized gas motions: \mbox{spirals} from the CND or \mbox{ellipses} around Sgr\,A$^*$. 
 Only the molecular structures A and B (cloudlet $\#$1) appear at the predicted positions and velocities of the Keplerian ellipse model. 
Instead, most HPVG and EPVG cloudlets   are distributed over a relatively extended area of the cavity (radius $<$15$''$ or 0.6~pc) and all show similarly high velocities and  broad line-widths  (\mbox{$\Delta$v$\simeq$25-40~km\,s$^{-1}$}, see Table~\ref{table:cloudlets} in the Appendix). 
Their line profiles are not doubled-peaked,
so they do not seem consistent with the expected emission from gas in rotating protoplanetary 
disks \mbox{\citep[e.g.,][]{Dutrey97}} or from  photoevaportaing disks \mbox{(proplyds)}, as previously suggested for other features  toward the cavity \citep[e.g.,][]{Yusef15}. 

Their similar line centroid velocity and broad line-widths, as well as  lack of associated bright ionized gas emission counterpart are puzzling. 
They may be compatible with the debris of a  tidally disrupted molecular cloud or clump  that fell into the region. 
The higher molecular gas velocities would
imply a  particularly fast molecular cloud, perhaps two, one for the HPVG and another one for the EPVG cloudlets, that fell from larger distances \citep{Oka11,Hsieh17} not so long ago; owing to the fast photo-evaporation timescales, got tidally disrupted in the cavity, and the densest leftovers are now resolved by ALMA as cloudlets. However, no physical process has been identified yet for scattering clouds onto orbits that have a low phase-space probability likelihood. Alternatively, the presence of  molecular cloudlets inside the cavity  might be accounted for instabilities in the inner rim of the CND that lead to fragmentation and infall from there \citep[e.g.,][]{Blank16}, through orbits that we have not modeled here. Such process may also allow faster replenishment of the molecular gas in the cavity.

In addition,  dissipative effects such as shocks in colliding streams, cloudlet-cloudlet collisions, and colliding stellar winds \citep[e.g.,][]{Calderon16}, or UV-induced dynamical effects produced by advancing photoionization fronts \cite[e.g.,][]{Goico16}, may trigger
enough gas compression and molecular gas formation. This may provide an explanation for structures A and B. Higher resolution observations of their proper motions are obviously needed to clarify their origin and fate.

\subsection{Molecular gas reservoir inside the cavity} 

From the CO~($J$=3-2) images  we have estimated the mass of the EPVG component inside the central cavity:  \mbox{$\sim$6-12~$M_{\odot}$} (range \mbox{$+$200$<$v$_{\rm LSR}$$<$$+$300~km\,s$^{-1}$}) plus \mbox{$\leq$45-190~$M_{\odot}$} in the HPVG range \mbox{($+$150$<$v$_{\rm LSR}$<$ +200$~km\,s$^{-1}$}). The low-$J$~CO emission from the latter velocity range, however, is affected  by lower-excitation CO extended emission from the background,  not related to  Sgr\,A$^*$ surroundings. Recent detection of high positive-velocity emission wings in more excited CO rotational lines (\mbox{up to $J$=10-9)} provides further evidence that molecular gas inside the cavity exists and is hot; contribution to the  \mbox{mid-$J$~CO}  emission from the  background  is likely negligible \citep{Goicoechea18}. These authors estimated a mass of $\sim$10-60~$M_{\odot}$
for this hot molecular gas component.

Overall, the inferred molecular mass inside the cavity is low, but much higher than that previously estimated from infrared H$_2$ line observations  \citep[$<$0.5~M$_{\odot}$,][]{Ciurlo16,Mills17}. Our derived molecular gas mass is in the order, or slightly below, of the estimated neutral atomic gas mass \citep[\mbox{$\sim$50-300~$M_{\odot}$},][]{Jackson93,Telesco96,Latvakoski99} and higher than the \mbox{$\sim$60~$M_{\odot}$} inferred in the ionized gas streamers \citep[e.g.,][]{Lo83,Jackson93,Irons12}. 
 As any dense photo-dissociation region (PDR) close to massive stars 
 \citep[e.g.,][]{Hollenbach97,Goico16}, 
  these different masses suggest that the cloudlets have a  stratified PDR 
   structure \mbox{(onion-like)}: with the infrared H$_2$ emission arising from the outermost and more \mbox{FUV-irradiated} layers, and the \mbox{low-$J$ $^{12}$CO} emission tracing the bulk of cloudlet mass. According to the low estimated masses, 
the cloudlets are not massive enough, individually, to be gravitationally bound  \mbox{($M_{\rm Cloudlet}\ll M_{\rm Virial}$)} so that it is unlikely that they will form new stars unless much material is accumulated, but see alternative scenarios in \citet{Yusef17}. Given the rapid photo-evaporation timescales, this process would have  to be quick enough to replenish this molecular gas reservoir in such a harsh, dynamic, and strong gravity environment.
        
Using the inferred HPVG and EPVG cloudlet masses and adopting representative velocities
of 150 and 250~km\,s$^{-1}$ for their motion, we estimate infall rates of
10$^{-2}$ and \mbox{10$^{-3}$~$M_{\odot}$\,yr$^{-1}$} respectively. This means
that the detected cloudlets inside the cavity are fresh molecular material that has accumulated only in the last $\lesssim$10$^4$\,yr. This is consistent with the photo-dissociation timescales of
cloudlets with $A_V$$\lesssim$4~mag. Hence, the detected extended and diffuse ($A_V$$\lesssim$0.5~mag)  molecular gas component has to be a very short-lived
component: continuously replenished by cloudlets' photo-evaporation and tidal disruption.  
 
\section{Summary and conclusions}\label{sec-summary}       

The Galactic center is about 100 times closer than any active galactic nucleus. Hence, it offers a unique
laboratory to study in exquisite detail the evolution of interstellar gas and massive stars in the strong gravitational potential of a SMBH and nuclear stellar cluster.
We have presented new images of the molecular gas  emission toward the central parsec of the Milky Way obtained with ALMA at 1$''$ angular resolution ($\sim$0.04~pc). Our results include the first unambiguous detection high velocity molecular cloudlets inside the cavity of the CND.

The ALMA images show that the southern lobe of the CND has a clumpy gas distribution, as traced by \mbox{H$^{13}$CN $J$=4-3} and  \mbox{SO~8$_9$-7$_8$ lines}.  Its motion requires a bundle of nonuniformly rotating streamers of slightly different inclinations. The inferred gas density peaks, molecular cores of several 10$^5$~cm$^{-3}$, are much lower than the local Roche density,  so we agree with previous works that the CND cores
are transient  \citep{Mills13,Smith14}, in other words, they will be stretched and disrupted instead of forming  gravitationally bound prestellar cores. However,  hydrodynamical simulations show that this conclusion does not immediately imply that the entire CND is a transient entity.
Instabilities within the CND, or in the interaction layer between the outflowing winds from massive stars  and the inner edge of the CND, can generate transient structures within a relatively stable CND  \mbox{\citep[e.g.,][]{Blank16}}.

Inside the cavity and in the inner rim of the CND, the molecular gas is irradiated by a strong FUV field produced by massive stars in the central cluster. We estimate
$G_{\rm 0,Cavity}$$\simeq$10$^{5.3}$ and $G_{\rm 0,CND}$$\simeq$10$^{4.3}$ respectively. The required  attenuation of the FUV radiation flux  implies $A_V$$\simeq$1-2 mag of local visual extinction within the cavity,
that is, it is not devoid of insterstellar gas and dust.
In particular, the detection of small spatial-scale $^{12}$CO gas structures close to Sgr\,A$^*$, and moving at high velocities,   confirms that the central cavity is not fully cleared of molecular gas.  It rather accepts a small fraction, at rates of \mbox{10$^{-2}$-10$^{-3}$~$M_{\odot}$\,yr$^{-1}$},
that  falls from the inner rim of the CND or  from  molecular  clouds that plunge from larger distances. In either case, molecular cloudlets
entering the cavity will be unstable against  tidal shear from the SMBH.
We have likely detected the products of this tidal fragmentation: the densest  fragments, the cloudlets, and a lower density and short-lived extended component. Evidence for such diffuse  intercloudlet gas is suggested  
by the $^{12}$CO images and also by the absorption spectrum toward Sgr~A$^*$ \mbox{synchrotron} emission.

 Except for two intriguing  structures (\mbox{A and B}
in \mbox{Figures~\ref{fig:Zoom}} and \ref{fig:ellipse_models}) we conclude that the velocities of the detected  cloudlets do not match those of the ionized 
streamers orbits. Instead, molecular cloudlets show similar velocity centroids and broad-line profiles  over a relatively large area of the cavity.
This may suggest that they were part of a more massive molecular cloud or clump, perhaps two, that fell to the inner region and got disrupted $\lesssim$10$^4$~yr ago. Yet, no process has been identified for scattering clouds onto such inner impact parameter orbits. Alternatively, the observed cloudlets may originate from instabilities in the inner rim of the CND that lead to fragmentation and infall from there.

It remains unclear whether this molecular
gas reservoir inside the cavity would ever grow enough to form a new generation of stars or if it would be photo-evaporated  and ionized by the strong stellar UV field in the region, completely disrupted by the strong gravitational tidal shears, or blown away by stellar winds from the central cluster.
While it is unlikely that the observed cloudlets will directly form new massive stars, their presence is a piece of the puzzle toward understanding the formation of stars close to SMBHs, the stability of CNDs,
and the influence of nuclear stellar clusters.

\begin{acknowledgements}

We thank our referee Mark Morris for useful comments and suggestions. We thank the ERC for support under grant \mbox{ERC-2013-Syg-610256-NANOCOSMOS}. We also thank Spanish MCIU for funding supports under grant AYA2017-85111-P.
JP and MG acknowledge by the Programme
National Physique et Chimie du Milieu Interstellaire (PCMI) of
CNRS/INSU with INC/INP co-funded by CEA and CNES.
 ALMA is a partnership of ESO (representing its member states), NSF (USA) and NINS (Japan), together with NRC (Canada), NSC and ASIAA (Taiwan), and KASI (Republic of Korea), in cooperation with the Republic of Chile. The Joint ALMA Observatory is operated by ESO, AUI/NRAO and NAOJ. 

\end{acknowledgements}


\bibliographystyle{aa}
\bibliography{references}

%-------------------------------------------------------------
%          For the appendices, table longer than a single page
%-------------------------------------------------------------

\begin{appendix} %First appendix

\section{ALMA images with and without total power single-dish maps}

%--------------------------------------------------------------------------------
\begin{figure*}[t]
\centering
\includegraphics[scale=0.22,angle=0]{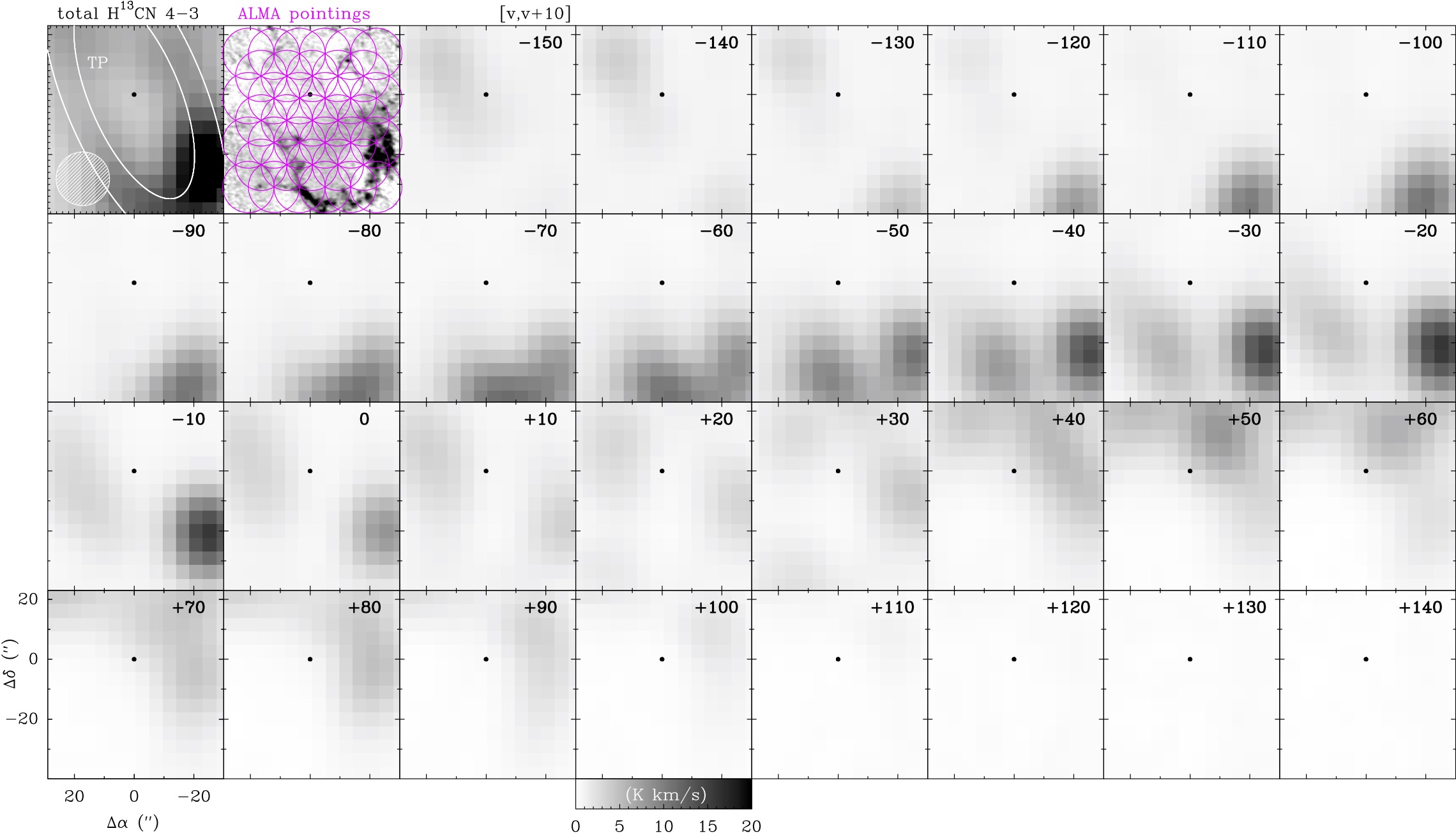}\\
\caption{Total power (single-dish) H$^{13}$CN~($J$=4-3) channel map observations at  $\sim$18$''$ angular resolution used as zero-spacings. The beam FWHM is shown by the hatched circle.
The second panel shows the merged ALMA+zero-spacing data as well as the ALMA interferometer array pointings.
}
\label{fig:h13cn_tp_appendix}
\end{figure*}
\begin{figure*}[t]
\centering
\includegraphics[scale=0.22,angle=0]{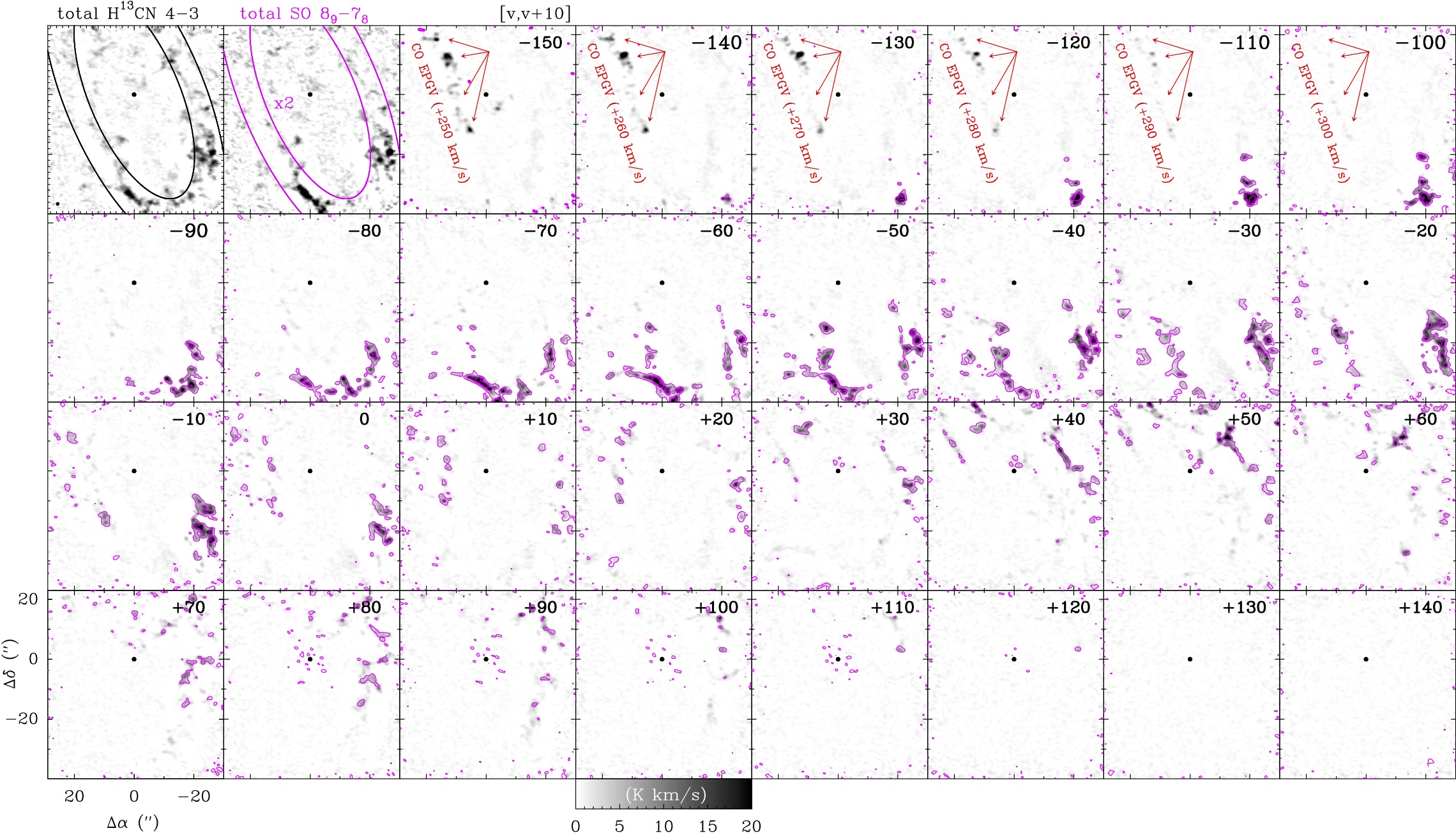}\\
\caption{Nearly identical spatial distribution of  H$^{13}$CN~($J$=4-3) and SO~(8$_9$-7$_8$) from ALMA interferometer data alone (zero-spacing data not merged). The~two top left panels show total integrated line intensity maps. The other maps show
H$^{13}$CN~($J$=4-3) (black images) and \mbox{SO~(8$_9$-7$_8$)} (magenta contours) channel maps  from v$_{\rm LSR}$=$-$150 to $+$150~km\,s$^{-1}$ in steps of 10~km\,s$^{-1}$. At all velocities, most H$^{13}$CN and SO features spatially match
(i.e., they belong to the same gas structure). The emission features seen at negative velocities without SO emission counterpart, however, do not arise from blue-shifted H$^{13}$CN emission but from red-shifted $^{12}$CO cloudlets. Sgr\,A$^*$~is at position (0$''$,0$''$). 
}
\label{fig:h13cn_so_only_alma_appendix}
\end{figure*}
\begin{figure*}[b]
\centering
\includegraphics[scale=0.22,angle=0]{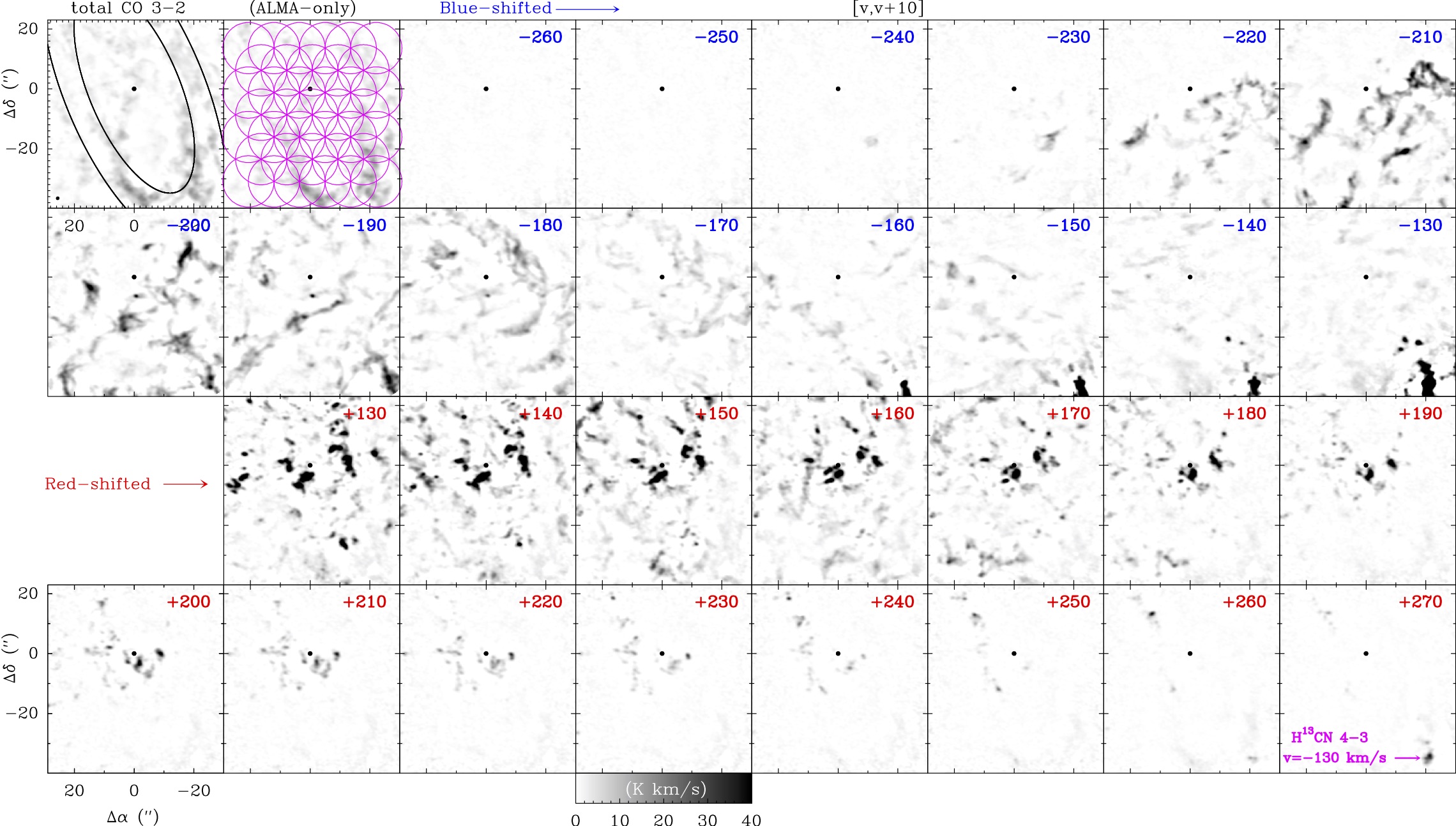}\\
\caption{ALMA (interferometer only, not including zero-spacing data) $^{12}$CO~($J$=3-2) channel emission maps
from v$_{\rm LSR}$=$-$260 to $-$120~km\,s$^{-1}$ (blue-shifted) and
from v$_{\rm LSR}$=$+$130 to $+$280~km\,s$^{-1}$ (red-shifted) in steps of 10~km\,s$^{-1}$.  
Sgr\,A$^*$, at position (0$''$,0$''$), is displayed with a black dot.}
\label{fig:co_alma-only_appendix}
\end{figure*}
\begin{figure*}[t]
\centering
\includegraphics[scale=0.22,angle=0]{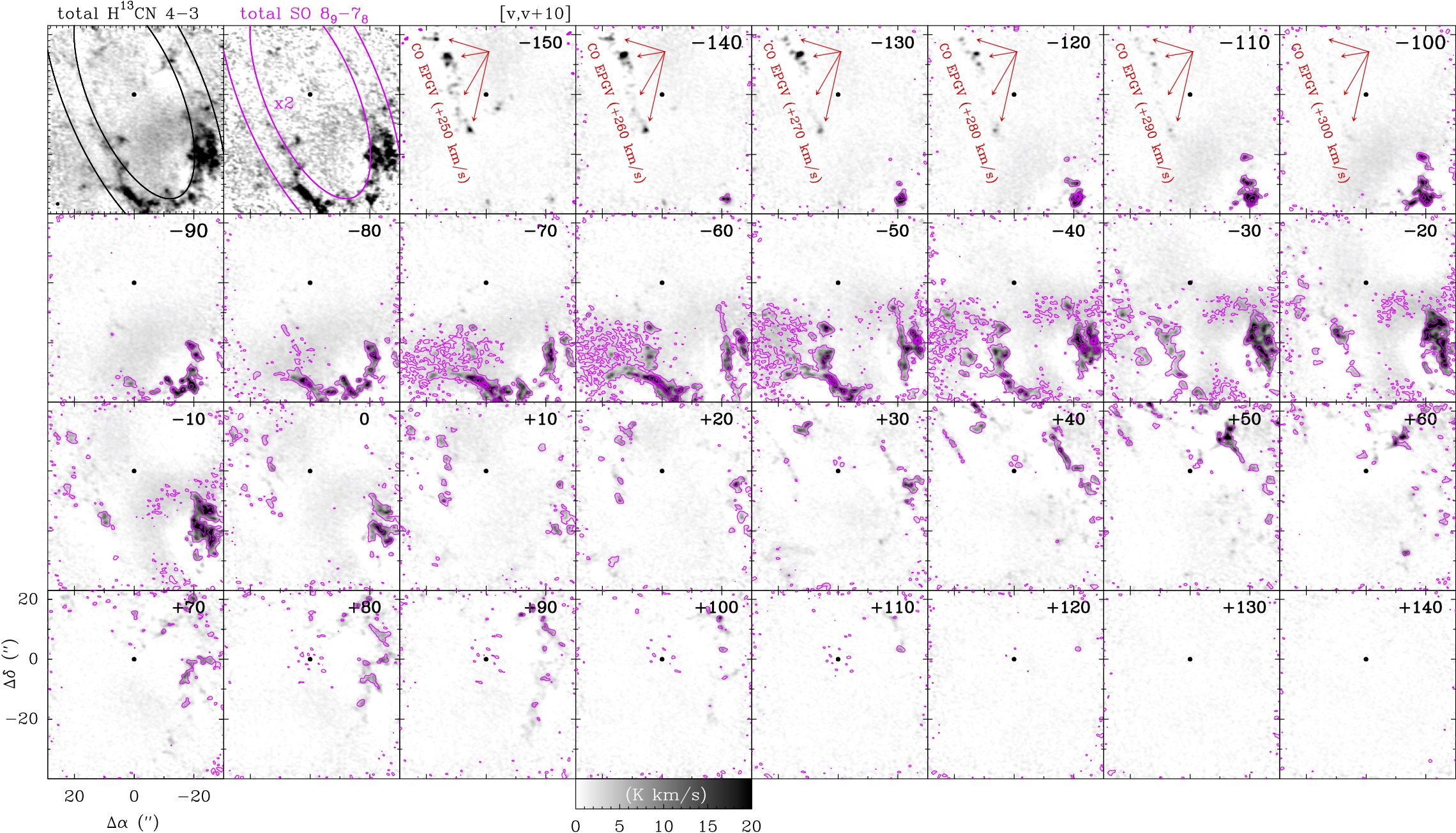}\\
\caption{Nearly identical spatial distribution of  H$^{13}$CN~($J$=4-3) and SO~(8$_9$-7$_8$) from ALMA+TP observations (merged zero-spacing data). The~two top left panels show total integrated line intensity maps. The other maps show
H$^{13}$CN~($J$=4-3) (black images) and \mbox{SO~(8$_9$-7$_8$)} (magenta contours) channel maps  from v$_{\rm LSR}$=$-$150 to $+$150~km\,s$^{-1}$ in steps of 10~km\,s$^{-1}$. At all velocities, most H$^{13}$CN and SO features spatially match
(i.e., they belong to the same gas structure). The emission features seen at negative velocities without SO emission counterpart, however, do not arise from blue-shifted H$^{13}$CN emission but from red-shifted $^{12}$CO cloudlets. Sgr\,A$^*$~is at position (0$''$,0$''$).}
\label{fig:h13cn_so_appendix}
\end{figure*}
\begin{figure*}[b]
\centering
\includegraphics[scale=0.22,angle=0]{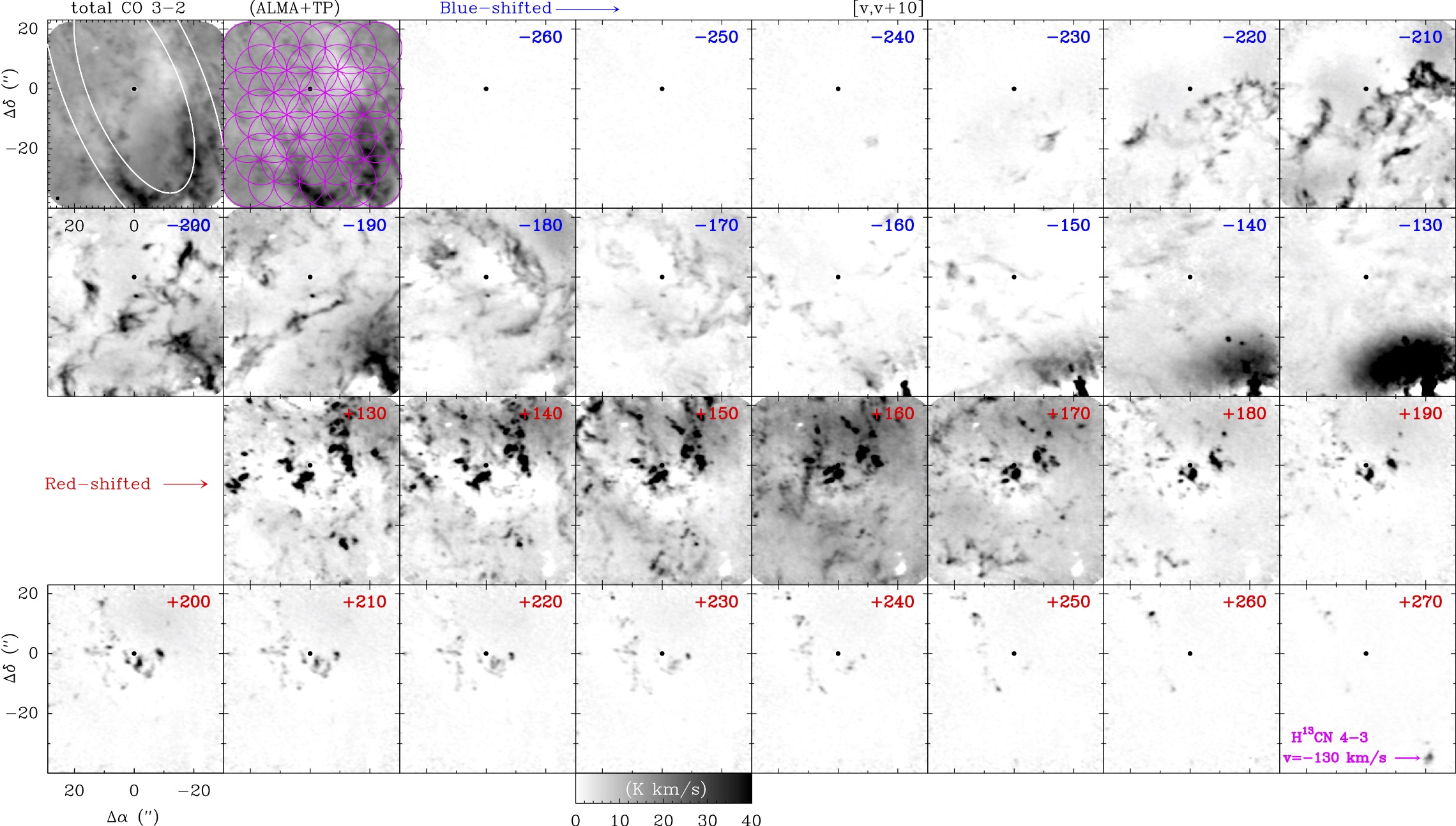}\\
\caption{ALMA+TP observations $^{12}$CO~($J$=3-2) channel emission maps
from v$_{\rm LSR}$=$-$260 to $-$120~km\,s$^{-1}$ (blue-shifted) and
from v$_{\rm LSR}$=$+$130 to $+$280~km\,s$^{-1}$ (red-shifted) in steps of 10~km\,s$^{-1}$.  
Sgr\,A$^*$, at position (0$''$,0$''$), is displayed with a black dot.}
\label{fig:co_alma-tp_appendix}
\end{figure*}

%--------------------------------------------------------------------------------
  Figure~\ref{fig:h13cn_tp_appendix} shows the deep total-power (TP) map obtained for H$^{13}$CN~($J$=4-3) and  merged with the ALMA \mbox{array} visibilities  to produce the final ALMA+TP images shown in the paper. %shown in Figure~\ref{fig:h13cn_so_appendix}.
\mbox{Figures~\ref{fig:h13cn_so_only_alma_appendix}} and
\ref{fig:co_alma-only_appendix}  show  \mbox{$^{12}$CO~($J$=3-2)}, \mbox{H$^{13}$CN~($J$=4-3)} line velocity channel maps 
obtained from ALMA interferometric data alone, without zero-spacing.
 The images show the intrinsically small spatial scale molecular emission 
 (HPVG and EPVG cloudlets).
\mbox{Figures~\ref{fig:h13cn_so_appendix} and \ref{fig:co_alma-tp_appendix}} show the final channel maps after adding the zero-spacing data (single-dish  maps) and revealing a more \mbox{extended emission}  filtered-out by the ALMA array-only data.

Figures~\ref{fig:h13cn_so_only_alma_appendix} and \ref{fig:h13cn_so_appendix}   demonstrate the nearly identical spatial distribution of the \mbox{H$^{13}$CN~($J$=4-3)}
and \mbox{SO 8$_9$-7$_8$} emission. The bright cloudlets appearing at negative velocities with respect to the \mbox{H$^{13}$CN~$J$=4-3} line frequency  (black images), but absent in the \mbox{SO (8$_9$-7$_8$)} images (magenta contours), are not produced by  H$^{13}$CN but by extreme positive \mbox{$^{12}$CO~($J$=3-2)} emission features.

\clearpage

%--------------------------------------------------------------------------------
%--------------------------------------------------------------------------------

%--------------------------------------------------------------------------------
%-------------------------------------------------------------------------------

%--------------------------------------------------------------------------------
%-------------------------------------------------------------------------------

%--------------------------------------------------------------------------------
%--------------------------------------------------------------------------------

\clearpage
\section{High and extreme positive-velocity gas cloudlet spectral fitting}

\begin{table*}[]
\caption{\label{table:cloudlets} Gaussian fit parameters of $^{12}$CO HPVG and EPVG cloudlets line emission (fit errors in parenthesis).} 
\centering
\begin{tabular}{ccccc}
\hline\hline
Cloudlet $\#$ & Offset$^a$ (arcsec) & Peak Velocity (km\,s$^{-1}$) & Line-width (km\,s$^{-1}$) & Line-peak (K) \\\hline
HPVG $\#$1 	& ($+$2.3,$-$5.0)    &	155.5 (0.1) & 37.9 (0.1)  &	13.6\\ 
HPVG $\#$2 	& ($-$1.0,$-$2.7)   &	166.3 (0.1) & 48.2 (0.1)  &	11\\
HPVG $\#$3 	& ($-$7.2,$+$3.1)    &	170.0 (0.4) & 28.9 (1.0)  &	10.2\\
HPVG $\#$4 	& ($+$9.3,$+$1.7)   & 	159.7 (0.1) & 31.4 (0.2)  &	10.1\\
HPVG $\#$5 	& ($-$6.1,$+$11.7) &	161.6 (0.9) & 28.0 (2.0)  &	6.8\\
HPVG $\#$6 	& ($-$11.3,$+$10.7)&	multi-comp.	& 	          & \\
EPVG $\#$7 	& ($-$8.9,$-$0.3)	& 204.9 ( 0.1) 	& 55.8 (1.1)  &	3.9\\
EPVG $\#$8 	& ($-$2.3,$-$3.0)   &	multi-comp.  &		      &\\
EPVG $\#$9 	& ($+$16.5,$+$18.7) &	242.1 (0.9) & 23.8 ( 2.2) &	3.7\\
EPVG $\#$10 & ($+$13.0,$+$13.0)	& 261.0 (0.5)   & 21.3 (1.2)  &	4.5\\
EPVG $\#$11 & ($+$5.6,$-$12.0)	& 254.6  (0.8)  & 28.6 (2.2)  &	2.1\\
EPVG $\#$12 & ($+$15.4,$+$1.1)	  & 219.5 (1.3)   & 30.5 (3.0)  &	1.6\\
EPVG $\#$13 & ($+$8.9,$+$5.5)	& 228.8 (1.1)   & 27.8 (2.4)  &	2.2\\
EPVG $\#$14 & ($-$4.2,$-$2.5) &	225.3 (1.1)	& 24.3 (3.1)  &	1.6\\
EPVG $\#$15 & ($+$3.7,$-$11.6)   &	212.4 (1.3) & 31.7 (3.4)  &	1.4\\
EPVG $\#$16 & ($+$9.0,0)	      & 222.6 ( 3.6)	& 40.0 (10.0) &	1.6\\\hline
\end{tabular}
\tablefoot{$^a$With respect to Sgr~A$^*$.}
%\textbf{References.}...}
\end{table*}

\end{appendix}

\end{document}